\documentclass[aps,prx,twocolumn,floatfix,longbibliography,superscriptaddress,nofootinbib,10pt]{revtex4-2}
\usepackage[utf8]{inputenc}
\usepackage[autostyle=try]{csquotes}
\usepackage{natbib}
\usepackage{graphicx}
\usepackage{comment}
\usepackage{subcaption}
\usepackage{xcolor}
\usepackage[tbtags]{amsmath}
\usepackage[colorlinks, linkcolor=blue]{hyperref}
\hypersetup{colorlinks,allcolors=black}
\usepackage{amssymb}
\usepackage{textcomp, gensymb}
\usepackage{float}
\usepackage{amsmath}
\usepackage{braket}
\usepackage{tabularx,graphicx}
\usepackage{epstopdf}
\usepackage{latexsym}
\usepackage{color, colortbl}
\usepackage{psfrag}
\usepackage{bbm}
\usepackage{bm}
\usepackage{titlesec}
\usepackage{dsfont}
\usepackage{feynmp}
\usepackage{slashed}
\usepackage{multirow}
\usepackage{siunitx}
\usepackage[normalem]{ulem}

\def \+ {\dagger}

\def \ve {\varepsilon}

\def \vp {\varphi}

\def \ve {\varepsilon}

\def \beq {\begin{eqnarray}}
\def \eeq {\end{eqnarray}}
\def \tn {\textnormal}

\def \nn {\nonumber}

\def \h {\mathcal{H}}
\def \htr {H_{\rm{transmon}}}
\def \hfl {H_{\rm{fluxonium}}}
\definecolor{applegreen}{rgb}{0.5, 0.8, 0}

\begin{document}

\title{Frozonium: Freezing Anharmonicity in Floquet Superconducting Circuits}
\author{Keiran Lewellen}
\author{Rohit Mukherjee}\thanks{These authors contributed equally to this work}
\author{Haoyu Guo}\thanks{These authors contributed equally to this work}
\author{Saswata Roy} 
\affiliation{Department of Physics, Cornell University, Ithaca NY 14853.}
\author{Valla Fatemi}
\affiliation{School of Applied and Engineering Physics, Cornell University, Ithaca NY 14853.}
\author{Debanjan Chowdhury}\email{debanjanchowdhury@cornell.edu}
\affiliation{Department of Physics, Cornell University, Ithaca NY 14853.}

\begin{abstract}

Floquet engineering is a powerful method that can be used to modify the properties of interacting many-body Hamiltonians via the application of periodic time-dependent drives. 
Here we consider the physics of an inductively shunted superconducting Josephson junction in the presence of Floquet drives in the fluxonium regime and beyond, which we dub the \textit{frozonium} artificial atom. 
We find that in the vicinity of special ratios of the drive amplitude and frequency, the many-body dynamics can be tuned to that of an effectively linear bosonic oscillator, with additional nonlinear corrections that are suppressed in higher powers of the drive frequency. By analyzing the inverse participation ratios between the time-evolved frozonium wavefunctions and the eigenbasis of a linear oscillator, we demonstrate the ability to achieve a novel dynamical control using a combination of numerical exact diagonalization and Floquet-Magnus expansion. We discuss the physics of \textit{resonances} between quasi-energy states induced by the drive, and ways to mitigate their effects. We also highlight the enhanced protection of frozonium against external sources of noise present in experimental setups. This work lays the foundation for future applications in quantum memory and bosonic quantum control using superconducting circuits.
\end{abstract}
\maketitle

\section{Introduction}
The past two decades have seen a series of remarkable breakthroughs in the field of circuit quantum electrodynamics \cite{blais2007quantum,PhysRevA.76.042319,RevModPhys.93.025005} --- the microwave-frequency quantum light-matter interactions between harmonic oscillators and superconducting artifical atoms.
At present, multiple commercially available quantum computing hardware platforms are based on superconducting qubits \cite{sqc1,sqc2,sqc3}, which leverage capacitively shunted and coupled Josephson junction circuits to create artificial atoms with individually-addressable quantum transitions \cite{PhysRevA.76.042319}. 
Through the manipulation and control of the occupation of these levels via microwave-frequency driving, quantum information can be processed to accomplish a variety of non-trivial tasks. 

The transmon regime of capacitively coupled Josephson junctions has been of great practical significance in recent times in the implementation of large arrays of qubits \cite{PhysRevA.76.042319,Google2024,Kim2023}. 
However, recent theoretical studies have shown that arrays of coupled transmons are generically prone to chaotic dynamics~\cite{berke2022transmon,borner2023classical,basilewitsch2024,resubref3,resubref4,resubref6}. 
Such chaos arises from the necessary non-linearity of superconducting qubits and shares many features with classical chaos in systems of coupled non-linear pendula \cite{borner2023classical,mukherjee2024}. 
Chaos poses a potential challenge for the further scalability of superconducting qubit arrays, as it can amplify errors and accelerate decoherence. 
At present, susceptibility to chaos is mitigated by the use of tunable couplers \cite{arute2019quantum} or via frequency detuning \cite{8936946}. 
However, both of these approaches introduce design and circuit complexity, potentially introducing new sources of decoherence; the viability of scalability to thousands of qubits remains presently unclear.

Alternative methods of addressing this susceptibility to chaos, and decoherence in general, are thus of significant interest~\cite{gyenis_moving_2021}. 
A potential avenue lies in Floquet engineering: using periodic time-dependent drives to tailor effective Hamiltonians with desirable properties. 
Supported by advances in Floquet analysis, it is an active area of study to this end~\cite{didier_ac_2019,petrescu_lifetime_2020,mundada_floquet-engineered_2020,huang_engineering_2021,venkatraman_static_2022,cohen_reminiscence_2023,petrescu_accurate_2023,xiao_diagrammatic_2024,thibodeau_floquet_2024,PhysRevA.109.042607,Blain_2025,Gandon_2022}. 
In this framework, carefully tuned drives could be used to control nonlinearities -- suppressing chaotic behavior when desired, while still allowing for non-linear regimes necessary for qubit gate operations. 
Ideally, such a Floquet qubit is also highly resistant to many dominant sources of external noise, as has been shown possible in a variety of Floquet qubits that were not intended to additionally suppress chaotic behavior \cite{didier_ac_2019,mundada_floquet-engineered_2020,huang_engineering_2021,thibodeau_floquet_2024,PhysRevA.109.042607}.

Previous work by some of us \cite{mukherjee2024} investigated this possibility in the context of driven transmons by making use of a method from Floquet theory known as \textit{dynamical freezing} \cite{PhysRevB.82.172402,KS2,PhysRevX.11.021008,sreemayee,guo2024dynamical,mukherjee2024,haldar2024dynamical}. 
We found that by tuning a high-frequency external drive to special \textit{freezing points} at fixed ratios of the drive amplitude to frequency, the non-linear Josephson energy of the qubits can be made to be highly suppressed in powers of inverse drive frequency.
This results in a strong suppression of chaos in coupled arrays, offering a promising potential route to stabilizing large-scale superconducting platforms. 
However, in the infinite frequency limit, the Hamiltonian loses its dependence on the superconducting phase of the transmon and thereby exhibits no quantum dynamics.
Furthermore, the offset charge was not considered in that work, and previous studies suggest it will result in a strong dephasing errors in the charge degree of freedom due to well-documented charge noise~\cite{koch_charge-insensitive_2007,kjaergaard_superconducting_2020,de_leon_materials_2021}. 
As such, a Floquet qubit that (1) allows for a tunable suppression of chaos while (2) simultaneously retaining its quantum dynamics for all drive frequencies and (3) allowing for strong suppression of many sources of decoherence remains desirable.

\begin{figure}[t]
    \includegraphics[width=1.0\linewidth]{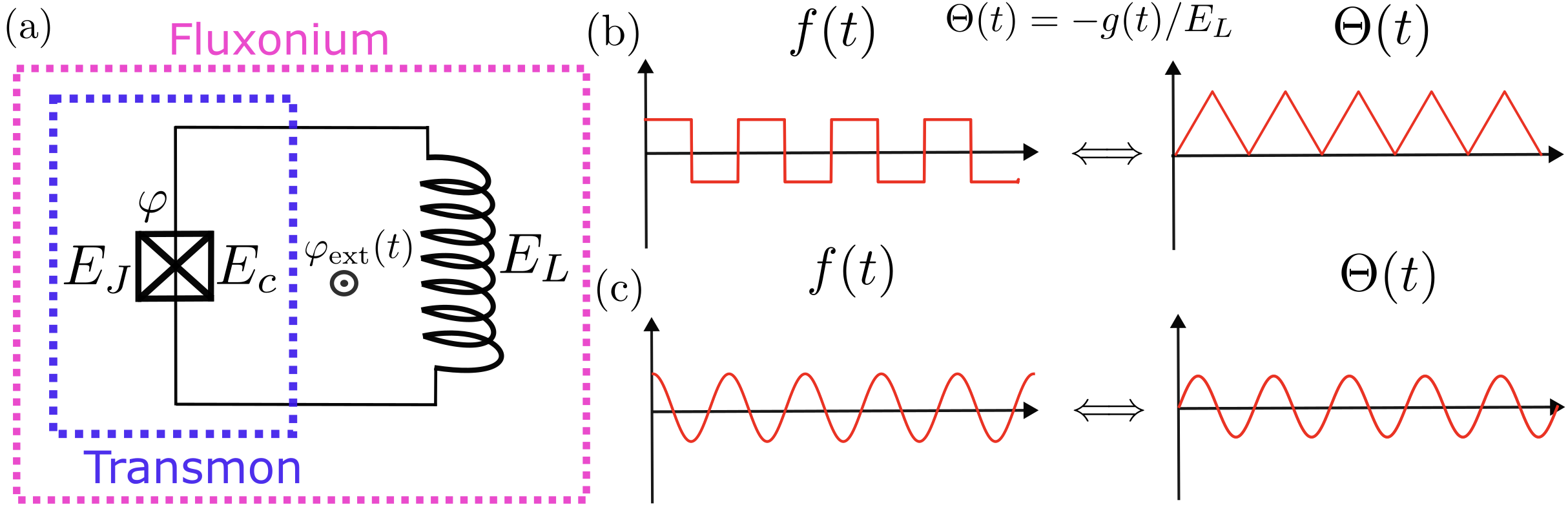}
\caption{\textbf{Figure 1: Schematic representation of the frozonium circuit and drive protocol.} (a) Schematic illustration of a transmon circuit consisting of a Josephson junction and capacitor (enclosed by blue dashed line), and a fluxonium circuit with an inductive shunt (enclosed by magenta dashed line). The (b) square-wave, and (c) cosine-wave Floquet drive, $f(t)$, in \eqref{Eq:fluxHam} controls the (b) triangle-wave, and (c) sine-wave Floquet drives, $\Theta(t)=\int_0^t dt'~f(t')$, respectively; see \eqref{eq:H_from_two_drives}.\label{fig:Schematic}
}
\end{figure}

With this in mind, the main focus of the present manuscript is on a modification of the previously considered driven transmon circuit with an inductive shunt, which we will find provides all of these benefits.
Our central novel result for the freezing points associated with the inductively shunted superconducting circuits will be their non-chaotic \textit{quantum} dynamics, which we dub as ``frozonium". 
Specifically, at the freezing point that is induced by a flux drive, the circuit is well-approximated as a harmonic oscillator at the leading order in inverse drive frequency. 
Therefore, the drives can interpolate the circuit between harmonic and strongly anharmonic configurations. 
Furthermore, the inductive shunt removes the offset charge, and, although these circuits admit an external flux which can be noisy, we show that the freezing points exhibit a largely flux-insensitive spectrum. 
Therefore, the frozonium can be utilized to increase robustness to external sources of noise and chaos, while additionally serving as a novel platform for both qubit and potentially bosonic formulations of quantum information processing.

To be concrete, consider the following driven transmon Hamiltonian considered in~\cite{mukherjee2024}, 
\begin{subequations}
\beq\label{Eq:totaHam}
\h(t) &=& \htr  + f(t)H_{\rm{drive};\widehat{n}},\\
\label{Hamiltonian1}
    \htr &=& -E_{\tn{J}}\cos{(\widehat{\vp})} + 4E_{c}(\widehat{n} - n_g)^{2}, \label{Hamiltonian2}\\
    H_{\rm{drive};\widehat{n}} &=& \widehat{n},
\eeq
\end{subequations}
where $\widehat{n}$ is the Cooper pair number operator, $\widehat{\varphi}$ is the superconducting phase satisfying the usual commutation relation $[\widehat{\varphi},\widehat{n}]=i$, and $n_g$ is the charge offset resulting from the compactness of $\widehat{\varphi}$ in the range ($-\pi,\pi]$~\cite{vool_introduction_2017}. Here
$E_{c}$ is the capacitive charging energy and $E_{\tn{J}}$ represents the Josephson energy; see Fig.~\ref{fig:Schematic}a. 
The Floquet drive, $f(t)$, has an amplitude, $A$, and period $T$, such that $f(t) = f(t+T)$ (i.e. with fundamental drive frequency, $\omega = 2\pi/T$). At special ratios of $A/\omega$, the periodic $\cos(\hat\varphi)$ term becomes strongly suppressed and the qubit dynamics are well described by an purely charge dependent effective Hamiltonian with small charge and phase dependent corrections suppressed in powers of $O(1/\omega)$~\cite{mukherjee2024}.

Let us now add an inductive shunt to the model of \eqref{Eq:totaHam},
\begin{subequations}
\beq\label{Eq:fluxHam}
\h(t) &=& \hfl  + f(t)H_{\rm{drive};\widehat{n}} + g(t) H_{\rm{drive};\widehat{\vp}},\nn \\ \\
\label{flux1}
    \hfl &=& \htr + \frac{E_{\tn{L}}}{2}\widehat{\vp}^2, \label{flux2}\\ 
    H_{\rm{drive};\widehat{\vp}} &=& \widehat{\vp},
\eeq
\end{subequations}
where the additional terms associated with $E_{\tn{L}}$ are due to the inductive energies, and we will find the second drive, $g(t)$, (stemming from a flux drive written in the appropriate irrotational gauge \cite{You_2019}) is necessary to extend the freezing phenomenology to this setting; see Fig.~\ref{fig:Schematic}a. 
Note that $\htr$ no longer has the offset charge parameter in the presence of the inductive shunt.
We leave the precise relationship between $f(t)$ and $g(t)$ unspecified for now, but as we show later, for a specific choice of drives they are effectively equivalent to a flux drive. 
Note that while we will continue to refer to the above circuit as ``fluxonium", the phenomenology we discuss below is not restricted to any specific hierarchy of energy scales among $E_{\tn{L}},~E_{\tn{J}}$ and $E_c$ \cite{manucharyan_fluxonium_2009}.
The new inductive term $\widehat{\vp}^2$  leads to two key differences of both practical and theoretical significance.

First, we will find that the effective Hamiltonian at freezing for the driven fluxonium is always quantum mechanical, even in the $\omega\rightarrow\infty$ limit. 
In that limit, the Hamiltonian is further both quantum mechanical and fully non-chaotic, unlike all previous examples of dynamical freezing \cite{PhysRevB.82.172402,KS2,PhysRevX.11.021008,sreemayee,guo2024dynamical,mukherjee2024,haldar2024dynamical}.
Perturbing away from the $\omega\rightarrow\infty$ limit down to realistic frequencies introduces additional quantum mechanical structure that has interesting implications for control of the residual nonlinearities. 

Second, the inductive shunt breaks the periodicity in $\widehat{\vp}$ of the Hamiltonian, effectively eliminating the static offset charge $n_g$ present in the conventional transmon~\cite{manucharyan_fluxonium_2009,koch_charging_2009}. 
Instead the formation of a closed loop in the circuit encompassing the inductive and Josephson elements results in flux noise through this loop being the primarily source of decoherence. 
Time-dependent fluctuations in this flux lead to a time dependence in the difference between the energies of the circuit. 
In particular, random variations in the energy difference between the computational states, $|\epsilon_1 - \epsilon_0|$ results in phase errors --- the build up of an unknown phase between the computational states~\cite{gyenis_moving_2021}.
The flux noise spectrum tends to be slow relative to the energy scales of the system~\cite{de_leon_materials_2021,wellstood_lowfrequency_1987,PhysRevLett.98.267003,anton_magnetic_2013,sendelbach_magnetism_2008,PRXQuantum.3.037001}. 
As we will demonstrate below, the driven fluxonium at freezing is significantly less sensitive to flux noise than the static circuit.

\section{Results}
\label{sec:results}
\subsection{Floquet-Magnus Expansion} 
\label{sec:FM}

To begin with, we describe the general intuition for the expected results by performing a Floquet-Magnus expansion; see Supplementary Material for additional details.
Moreover, this will also help clarify the required relationship between the two drive functions, $f(t)$ and $g(t)$, that can enable the novel control of the induced nonlinearities. In order to obtain the quasi-energy spectrum and late-time behavior associated with the model in \eqref{Eq:fluxHam}, we first perform a time-dependent unitary transformation to a co-moving reference frame, $ W(t)=e^{-i\Theta(t) H_{\rm{drive};\widehat{n}}}$, where $\Theta(t) = \int_0^t dt' f(t')$. The transformed Hamiltonian is then,
\begin{subequations}
\beq
    \mathcal{H}_\text{mov}(t) &=& W^\dagger(t)[\mathcal{H}(t) -\partial_t]W(t)\\
    &=& W^\dagger(t)[H_\text{fluxonium} + g(t) H_{\rm{drive};\widehat{\vp}}]W(t).
    \eeq
\end{subequations}
This can be simplified further using the Baker-Campbell-Hausdorff (BCH) formula to yield,
\begin{align}
    \mathcal{H}_\text{mov}(t) = \ &\htr[\widehat{n}, \widehat{\vp} - \Theta(t)]  + \frac{E_{L}}{2}\widehat{\vp}^2 \nonumber\\
    &+ \Theta(t)E_{L} \widehat{\vp} + g(t)\widehat{\vp}.
\end{align}
If we now choose a specific relationship between the drives such that $ g(t) = -E_{\tn{L}}\Theta(t)$, the last two terms above cancel mutually, and we arrive at the significantly simplified Hamiltonian,
\begin{align}
    \mathcal{H}_\text{mov}(t) = \ &\htr[\widehat{n}, \widehat{\vp} - \Theta(t)]  + \frac{E_{L}}{2}\widehat{\vp}^2.\label{eq:H_from_two_drives}
\end{align}
Allowing for the presence of an external flux, $\vp_\text{ext}$, through the inductive-Josephson loop gives, 
\begin{subequations}
\beq
    \mathcal{H}_{\text{mov}}(t) &=& H_\text{quad} - E_{\tn{J}}\cos(\widehat\vp +\vp_\text{ext} - \Theta(t))\label{eq:H_mov},\\
    H_\text{quad} &=& 4E_c\widehat{n}^2 +\frac{E_{\tn{L}}}{2}\widehat{\vp}^2.\label{eq:quad}
    \eeq
\end{subequations}
When making reference to any analytical discussions of the problem, this is the Hamiltonian we work with in the remainder of the manuscript. 
Unless otherwise stated, we fix the static flux $\vp_{\rm{ext}}=\pi$ in the remainder of the manuscript.

For the remainder of this manuscript, we consider two forms of the periodic Floquet drive, $\Theta(t)$: the triangle-wave drive and the sine-wave drive; see Fig.~\ref{fig:Schematic}b,c. Specifically,
\begin{subequations}
\beq
    \Theta_{\text{tw}}(t) &=& \alpha\cdot2\pi\left|\frac{\omega}{2\pi}t - \left\lfloor\frac{\omega}{2 \pi}t + \frac{1}{2}\right\rfloor\right|,\label{eq:tw_drive}\\
    \Theta_\text{sin}(t) &=& \alpha\sin(\omega t),\label{eq:sin_drive}
\eeq
\end{subequations}  
where $\lfloor\cdot\rfloor$ indicates taking the integer part of a real number (see Fig.~\ref{fig:Schematic}c), 
$\alpha$ is the phase amplitude of the drive, and $\omega$ its frequency which we will take to be in the gigahertz (GHz) range. (Note that while $\Theta_\text{tw}(t)$ has a net DC offset, this is notational, and the offset can be absorbed into the definition of $\vp_\text{ext}$.) 
The original drives are thus given by
\begin{subequations}
\beq
    f_\text{tw}(t) = A\ \text{sgn}[\sin(\omega t)],\ g_\text{tw}(t) = -E_{\tn{L}}\Theta_\text{tw}(t),\label{eq:sq_transmon} \\
    f_\text{sin}(t) = A\cos(\omega t),\ g_\text{sin}(t) = -E_{\tn{L}}\Theta_\text{sin}(t),
\eeq 
\end{subequations}
where $A$ is the amplitude of the $\widehat{n}$ drive, and $\alpha = A/\omega$, respectively. 
We focus primarily on results for the triangle-wave drive in the main text for which the phenomenology of frozonium is more pronounced, and leave a discussion of the results for the sine drive for Supplementary material. 
Furthermore, unless otherwise noted, we will work with $E_c/h = 0.33$ GHz, $E_{\tn{L}}/h = 1$ GHz, and $E_{\tn{J}}/h = 4$ GHz ~\cite{fluxonium1}.

Starting with the Hamiltonian in \eqref{eq:H_mov}, the leading order term of the Magnus expansion is given by,
\begin{align}
    \mathcal{H}^0_\text{eff} = \frac{1}{T}\int_0^T dt~ \mathcal{H}_\text{mov}(t),
\end{align} 
where depending on the form of the periodic drive, we obtain,
\begin{subequations}
\beq
    \mathcal{H}^0_\text{eff;tw} &=& H_\text{quad} - \frac{E_{\tn{J}}}{\pi\alpha}(\sin(\widehat{\vp}) + \sin(\pi\alpha - \widehat{\vp})) + \cdots, \label{eq:mag_zero_tw}\nn\\ \\
    \mathcal{H}^0_\text{eff;sin} &=& H_\text{quad} -E_{\tn{J}} J_0(\alpha)\cos(\widehat{\vp}) + \cdots.\label{eq:mag_zero_sin}
\eeq
\end{subequations}
Here $J_0(...)$ is the Bessel function of first kind, and the $\cdots$ represent corrections suppressed in $O(1/\omega)$. 
Thus, the leading order nonlinearity vanishes at certain fixed parameter values:
\begin{subequations}
 \beq
     \alpha &\approx& 2n,\ n\in\mathbb{Z}~~(\rm{Triangle ~wave})\label{eq:tw_freezing_points}\\
     \alpha &\approx& \text{zeros of $J_0(...)$}~~(\rm{Sine ~wave})~.
 \eeq
 \end{subequations}
 As a result, the effective Hamiltonian for the driven fluxonium represents a linear \textit{quantum} harmonic oscillator given by \eqref{eq:quad}, with all of the higher-order nonlinearities suppressed in $O(1/\omega)$ (see Supplementary material for additional details). 
 As noted previously, this is distinct from {\it all} previous discussions of dynamical freezing (including the driven transmons \cite{mukherjee2024}), where the effective Hamiltonian in the $\omega\rightarrow\infty$ limit is purely classical. 
 Therefore, this relatively straightforward analysis already suggests that by varying both $\alpha$ and $\omega$, we can tune the relative strength of nonlinearity associated with the dynamics of the effective Hamiltonian in the vicinity of the freezing points. 
 Next we will turn to a direct numerical analysis of the driven fluxonium.

\subsection{Numerical Exact Diagonalization}\label{sec:ED}

\begin{figure*}[t]
    \includegraphics[width=1.0\linewidth]{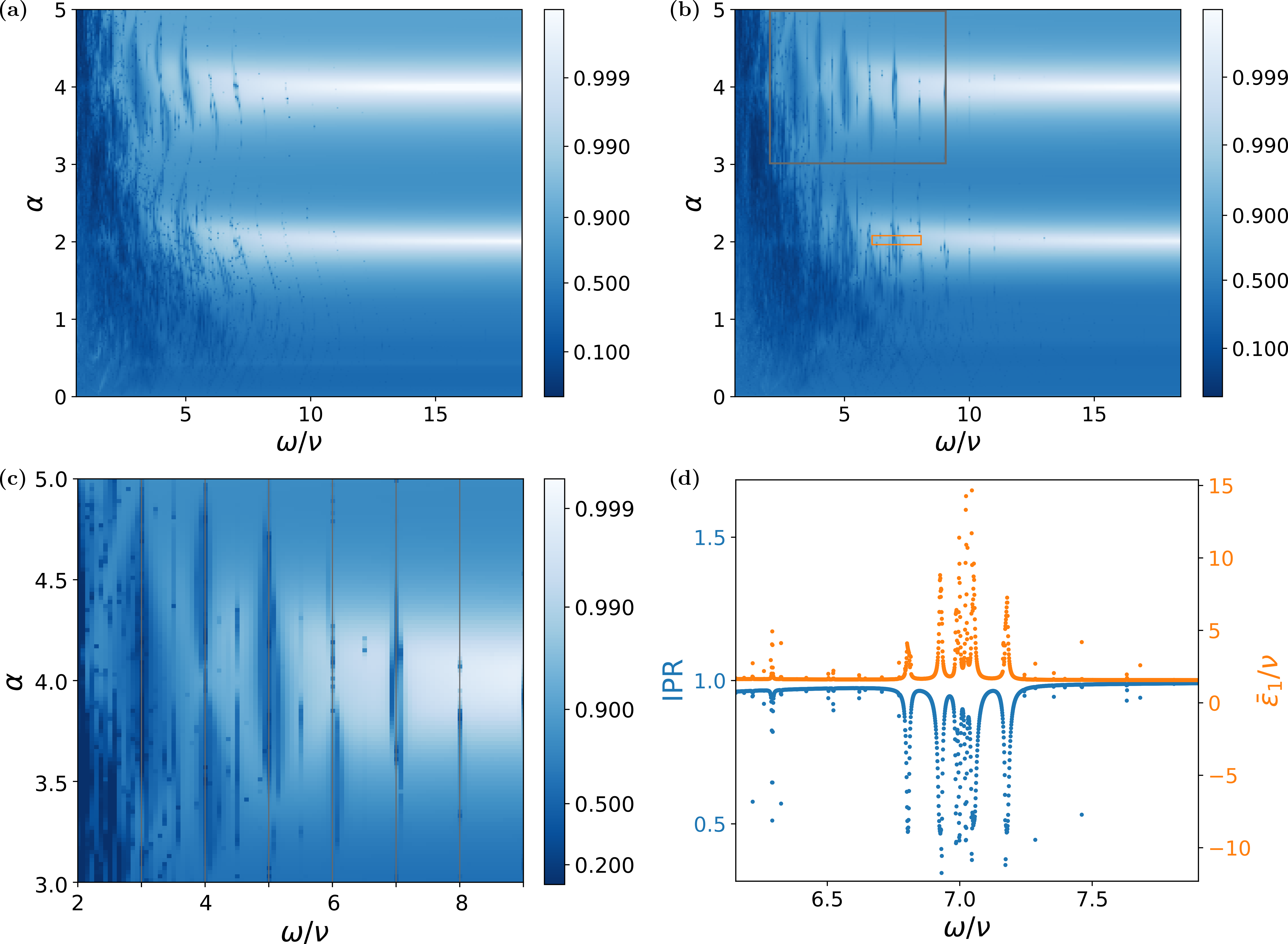}
\caption{\textbf{Figure 2: Freezing points and resonances in frozonium.} The IPR (\eqref{eq:IPR}) as a function of $\alpha$ and $\omega/\nu$ for a fluxonium driven via $\Theta_{\tn{tw}}(t)$ (\eqref{eq:tw_drive}) with $\varphi_{\tn{ext}}=\pi$ time-averaged over $N_F=10,000$ cycles. Here $\nu$ is the fundamental frequency (given in Eq.~\ref{eq:osc_freq}) of the oscillator defined in \eqref{eq:quad}. $\ket{\Psi(0)}$ is initialized to the (a) ground state, and (b) first excited state of $H_{\tn{quad}}$ in \eqref{eq:quad}. The white horizontal bands with $\text{IPR}>0.99$ centered around $\alpha = 2,~4$ are associated with freezing region (see \eqref{eq:tw_freezing_points}). (c) Zoom-in from panel (b) [grey rectangle] showing vertical streaks of $\text{IPR}\ll1$ embedded in a region with $\text{IPR}\approx1$. The color-scale in (a)-(c) is chosen to have an arctanh form, normalized to a domain, $(1/d_H, 1)$, with $d_H=70$. We adopt the Hamiltonian parameters \( E_c/h = 0.33 \, \text{GHz} \), \( E_{\tn{L}}/h = 1 \, \text{GHz} \), and \( E_{\tn{J}}/h = 4 \, \text{GHz} \), motivated by Ref.~\cite{fluxonium1} (d) $\text{IPR}$ at fixed $\alpha$ as a function of $\omega/\nu$ (blue curve) from panel (b) [orange rectangle]. The frequencies with a drop in $\text{IPR}$ correlate with peaks in $\bar{\ve}_k$ for the first excited state defined in \eqref{eq:Period_avg_E}.\label{fig:IPR}
}
\end{figure*}

Let us now simulate numerically the system of a single fluxonium driven with a triangle-wave drive, $\Theta(t)$; see Methods for additional details. Based on our discussion in Sec.~\ref{sec:FM}, we expect the effective Hamiltonian near freezing (see \eqref{eq:tw_freezing_points}) to be well described by the quadratic Hamiltonian in \eqref{eq:quad}. Our goal is to diagnose the extent to which the numerically exact eigenstates for the full driven fluxonium overlap with the eigenstates of the harmonic oscillator described by $H_{\rm{quad}}$ in \eqref{eq:quad}. To quantify this, we first choose an eigenstate of $H_{\rm{quad}}$ as the initial state, and time-evolve it with the driven Hamiltonian, ${\cal{H}}_{\rm{mov}}(t)$ in \eqref{eq:H_mov} for $N_F$ Floquet cycles. We compute the inverse participation ratio (IPR) at the end of the time evolution, which is defined as
\begin{subequations}
\beq
    \text{IPR} &=& \sum_{i=1}^{d_H}|\braket{\Phi_i|\Psi(t)}|^4\label{eq:IPR},~\tn{where}\\
    \ket{\Psi(t)} &=& \mathcal{T}\exp\left(-\int_0^{t}dt'i{\cal{H}}_{\tn{mov}}(t')\right)\ket{\Psi(0)},
\eeq
\end{subequations}
and $\mathcal{T}$ denotes time-ordering. Here $\{\ket{\Phi_i}\}$ represents a complete eigenbasis associated with the quantum harmonic oscillator in \eqref{eq:quad}. Note that while the local Hilbert-space is formally unbounded, for our numerical computations we truncate to a relatively large low-energy subspace with $d_H=70$. For evaluating the IPR, we choose $N_F=10,000$ cycles.

At time $t=0$, the IPR starts out with a value of $1$. For a given value of $\alpha$ and $\omega$, if the effective Hamiltonian for the driven system is well approximated by $H_{\rm{quad}}$, we expect $\text{IPR}\approx1$. On the other hand, if the system ``equilibrates" to a featureless ``infinite-temperature" state as a result of the time-evolution under the full nonlinear theory, we expect $\text{IPR} \sim O(1/d_H)$, where $d_H$ is the Hilbert space dimension. Generically, away from the freezing points, the IPR will assume a value that is different from these two limiting cases. The actual value is a function of the drive frequency, proximity to freezing, and $N_F$. This indicates that the effective Hamiltonian is not well approximated by $H_{\rm{quad}}$, and the inevitable approach to thermalization is then controlled by pre-thermal physics. Importantly, even the freezing points can eventually thermalize depending on the driving protocol \cite{mukherjee2024instanton}, but the associated timescales can be exponentially long in the drive-frequency, and therefore not of any practical concern. 

In Fig.~\ref{fig:IPR} we analyze the stroboscopically averaged $\text{IPR}$ over $N_F(=10,000)$ Floquet cycles as a function of $\alpha$ and $\omega/\nu$. Here $\nu$ is the fundamental energy of the oscillator described by $H_\text{quad}$ given by 
\begin{align}
    \nu/h = \sqrt{8E_CE_{\tn{L}}}/h\approx 1.63 \text{ GHz}.\label{eq:osc_freq}
\end{align}
We initialize  $\ket{\Psi(0)}$ to the ground state (Fig.~\ref{fig:IPR} (a)) and the first excited state (Fig.~\ref{fig:IPR} (b)) of $H_{\rm{quad}}$ respectively. As expected from \eqref{eq:tw_freezing_points}, we find clear evidence of horizontal ``bands" (white regions in Figs.~\ref{fig:IPR}a, b) centered around $\alpha = 2, 4, ...$ where the $\text{IPR}\gtrsim0.99$. Furthermore, the IPR is closer to $1$ above a threshold frequency set by the intrinsic single-particle energy scales, as noted previously in the context of freezing in Ref.~\cite{mukherjee2024}. We have also noticed that the width of the vertical region at low $\omega$ with a small IPR increases for $\ket{\Psi(0)}$ with higher energy. In order to recover an $\text{IPR}\rightarrow1$ for such states, the drive frequency needs to be correspondingly larger.  

We discuss the evolution of the IPR as a function of $N_F$ as well as the IPR for the sine drive-wave drive in the Supplementary material.

Let us now turn to one of the most striking features of our results in Fig.~\ref{fig:IPR}a, b, associated with the narrow vertical ``streaks" with a suppressed value of $\text{IPR}\ll 1$ embedded in a region near freezing with $\text{IPR}\rightarrow1$. In Fig.~\ref{fig:IPR}c, we show a zoomed-in region of the data in Fig.~\ref{fig:IPR}b centered around $\alpha=4$ that shows these streaks clearly. As we shall now discuss, these regions are associated with pairs of states whose quasi-energies differ exactly by the drive frequency $\omega$, resulting in a {\it resonance} phenomenon. At resonance, the pair of states are strongly hybridized due to the Floquet drive, resulting in spike-like features in the Floquet quasi-energies.  
To demonstrate the correlation between IPR drop and spike features of quasi-energies, let us now introduce a period-averaged Floquet quasi-energy over $0\leq t\leq T$ ~\cite{PhysRevE.82.021114,cohen_reminiscence_2023},
\begin{subequations}
\beq
    \bar{\ve}_k &=& \frac{1}{T}\int_0^T dt \bra{\phi_k(t)} {\cal{H}}_{\rm{mov}}(t)\ket{\phi_k(t)}\label{eq:Period_avg_E},\\
     \ket{\phi_k(t)} &=& \mathcal{T}\exp\left(-\int_0^{t}dt'i{\cal{H}}_{\tn{mov}}(t')\right)\ket{\phi_k(0)}.
\eeq 
\end{subequations}
Here $\ket{\phi_k(t)}$ are the Floquet eigenstates of $H_{\tn{mov}}(t)$, with $\ket{\phi_k(T)} = \ket{\phi_k(0)}$. Note that $\bar{\ve}_k$ as defined above, is not folded modulo $\omega$ and it serves as a proxy for studying the spectrum of the Floquet Hamiltonian. By analyzing the $\text{IPR}$ at a fixed $\alpha$ as a function of $\omega$ (see orange rectangle in Fig.~\ref{fig:IPR}b), we find that the precipitous drop away from $\text{IPR}\approx1$ coincides with strong peaks in $\bar{\ve}_k$; see Fig.~\ref{fig:IPR}d. Relatedly when $\text{IPR}\approx1$ away from the streak-like regions, $\bar{\ve}_k$ remains featureless.

\begin{figure}
\centering
    \includegraphics[width=\linewidth]{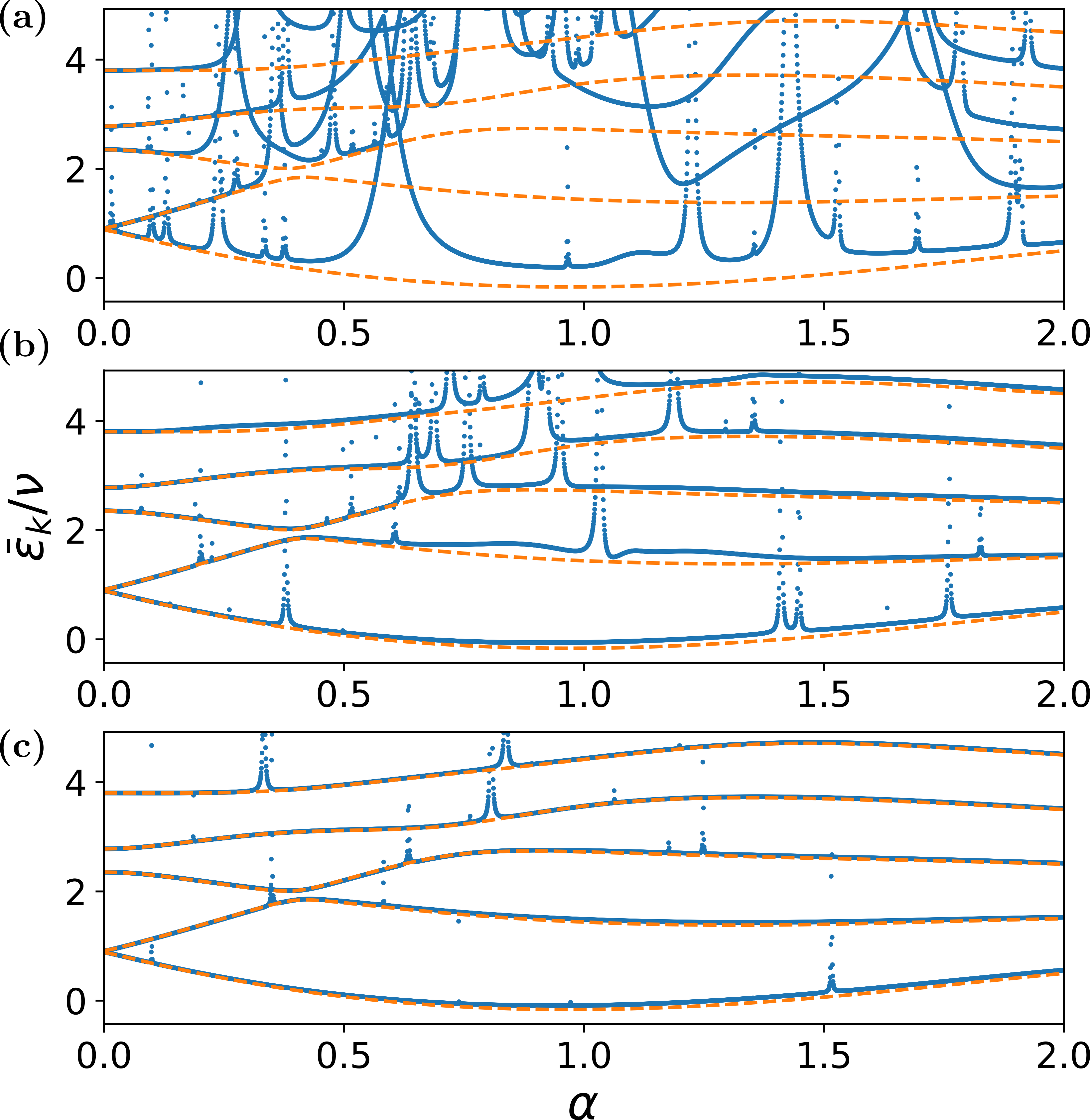}
    \caption{\textbf{Figure 3: Dependence of frozonium quasi-energies on drive parameters.} The period averaged quasi-energies, $\bar{\ve}_k$ (\eqref{eq:Period_avg_E}) as a function of $\alpha$ for a driven fluxonium at (a) $\omega/\nu = 6.15$, (b) $\omega/\nu = 9.23$, and (c) $\omega/\nu=12.31$, respectively. The result is normalized by the fluxonium oscillator energy $\nu$ (\eqref{eq:osc_freq}). All other parameters are as in Fig.~\ref{fig:IPR}.   Blue data points are obtained from numerical exact diagonalization. Orange dashed lines are obtained using the zeroth-order Floquet-Magnus expansion (\eqref{eq:mag_zero_tw}). 
}\label{fig:E_vs_alpha}
\end{figure}

To better characterize the origin of the resonances introduced above, we plot the five lowest period averaged quasi-energies (blue dots) for the driven fluxonium in Fig.~\ref{fig:E_vs_alpha} as a function of $\alpha$, and for three different values of increasing $\omega$. 
The $\ket{\phi_k(t)}$ in Eq.~\ref{eq:Period_avg_E} used to evaluate these are obtained from the numerical exact diagonalization. 
In the figure, we also show $\bar{\ve}_k^\text{FM}$  (orange dashed line) for the same five levels based on the zeroth-order Floquet-Magnus expansion (\eqref{eq:mag_zero_tw}). 
As anticipated, we find better agreement between the two sets of results with increasing drive frequency. 
Additionally, the number of resonances and the associated singular features in $\bar{\ve}_k$ decrease with increasing frequency. 
With increasing frequency, we also notice that the quasi-energy level-spacing interpolates more smoothly between the anharmonic regime ($\alpha = 0$) to the harmonic regime near freezing ($\alpha = 2$).

\begin{figure}
\centering
    \includegraphics[width=\linewidth]{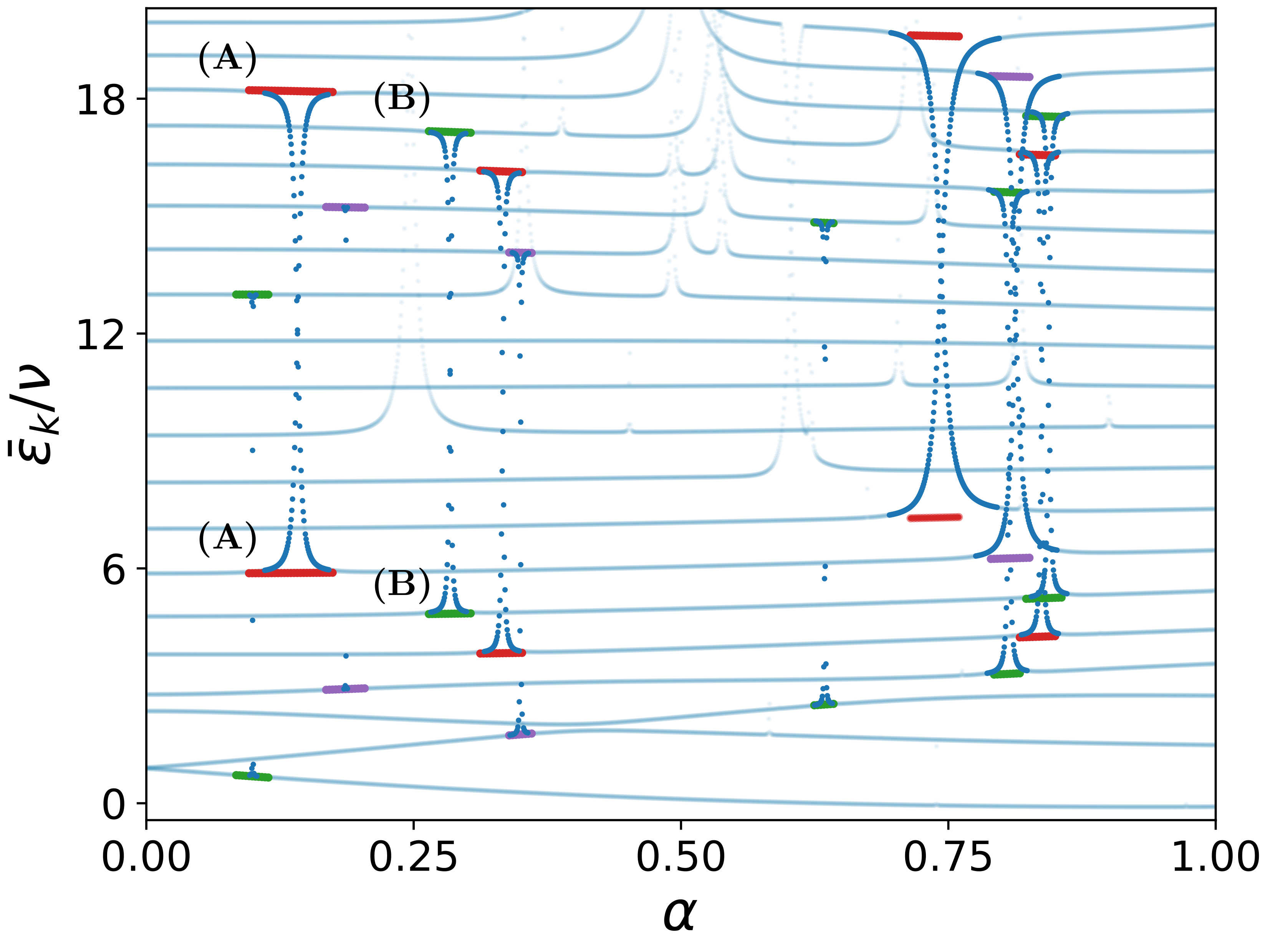}
    \caption{\textbf{Figure 4: Prediction of resonance points in frozonium.} The period averaged quasi-energies, $\bar{\ve}_k$, as a function of $\alpha$ for a driven fluxonium at $\omega/\nu =  12.31$. The result is normalized by the fluxonium oscillator energy $\nu$ (\eqref{eq:osc_freq}). All other parameters are as in Fig.~\ref{fig:IPR}.  
    The locations for the resonances determined from the numerically evaluated averaged quasi-energies (blue) agree well with those obtained from the zeroth order Floquet-Magnus results, which are marked by pairs of red, green, and purple bars.  The latter is obtained using the period-averaged quasi-energy $\bar{\ve}_i^\text{FM}$ from the Floquet Magnus expansion and the condition $|\bar{\ve}_i^{\text{FM}} - \bar{\ve}_j^{\text{FM}}| \in (\omega-\Delta\omega,\omega+\Delta\omega)$, where  $\omega=12.31\nu$ and we choose a small $\Delta\omega=0.03\nu$.
   As an example, we mark two such pairs of states as (A) and (B).
     Note that the resonances are only plotted for partners that lie within the twenty quasi-energies.}\label{fig:mag_resonance}

\end{figure}

Thus far, we have not shown the exact pairs of quasi-energy levels that control these resonances. 
To track these down, in Fig.~\ref{fig:mag_resonance} we have plotted the lowest twenty period averaged quasi-energies along with the predictions of the zeroth-order Floquet-Magnus expansion for the drive frequency in Fig.~\ref{fig:E_vs_alpha}c. 
We have scanned for all pairs of period-averaged quasi-energies $\bar{\ve}_i^\text{FM}$ obtained from the zeroth-order Floquet-Magnus where $|\bar{\ve}_i^\text{FM} - \bar{\ve}_j^\text{FM}| \in (\omega-\Delta\omega, \omega + \Delta\omega)$ (with $\Delta\omega = 0.03\nu \ll\omega$) as a function of $\alpha$. 
We have marked these pairs of quasi-energy levels [as (A), (B) and with colored bars] in Fig.~\ref{fig:mag_resonance}. 
At these values of $\alpha$, we expect a strong hybridization between Floquet eigenstates due to resonances. These resonances are between states separated by $\omega$ in $\bar{\ve}_k$, and are known to be related to heating of the driven system and are also notable in the context of qubit read-out protocols \cite{PhysRevLett.132.100401,PhysRevApplied.18.034031,PhysRevX.14.041023,féchant2025offsetchargedependencemeasurementinduced,cohen_reminiscence_2023}.
We find a one-to-one correspondence between the resonances in the period averaged Floquet quasi-energies obtained from the zeroth-order Floquet-Magnus expansion and the direct numerical results. To summarize, the Floquet-Magnus expansion can provide a good approximation for the smooth features of the period-averaged Floquet quasi-energies, but fails to capture the spike-like features associated with resonances. However, the location of the resonances can still be approximately determined from the Floquet-Magnus result.

An important limiting case of these resonances occurs near freezing points where the effective Hamiltonian for the driven fluxonium is well approximated by $H_\text{quad}$. At these points the energy level-spacing between the eigenstates becomes approximately uniform and is given by $\nu$ (see Eq.~\ref{eq:osc_freq}). Strong resonances are thus expected to occur for drive frequencies satisfying $\omega/\nu\in\mathbb{Z}$. These resonances can be seen clearly in Fig.~\ref{fig:IPR}c in which we plot the stroboscopically averaged $\text{IPR}$ (w.r.t. first excited state of $H_\text{quad}$) of a single driven fluxonium near freezing as a function of $\alpha$ and $\omega/\nu$. At integer values of $\omega/\nu$, we observe the sharp decreases in IPR indicative of resonances. Additionally, weaker resonances are marked by the suppression in $\text{IPR}$ at the half-integer ratios of $\omega/\nu$. At these points, there exist eigenenergies separated by a gap of approximately $2\omega$, resulting in the observation of additional resonances. It is worth pointing out that for the sine-wave drive at freezing, the leading nonlinear correction to the effective Hamiltonian beyond $H_{\rm{quad}}$ scales as $\sim \hat{n}\cos\hat{\vp}/\omega$. By choosing $\ket{\Psi(0)}$ to be an eigenstate of this nonlinear oscillator, we find an increase in the magnitude of the IPR compared to an eigenstate of $H_{\rm{quad}}$.

\subsection{Robustness to Variations in External Flux} 
\label{sec:noise}

\begin{figure*}
    \includegraphics[width=1.0\linewidth]{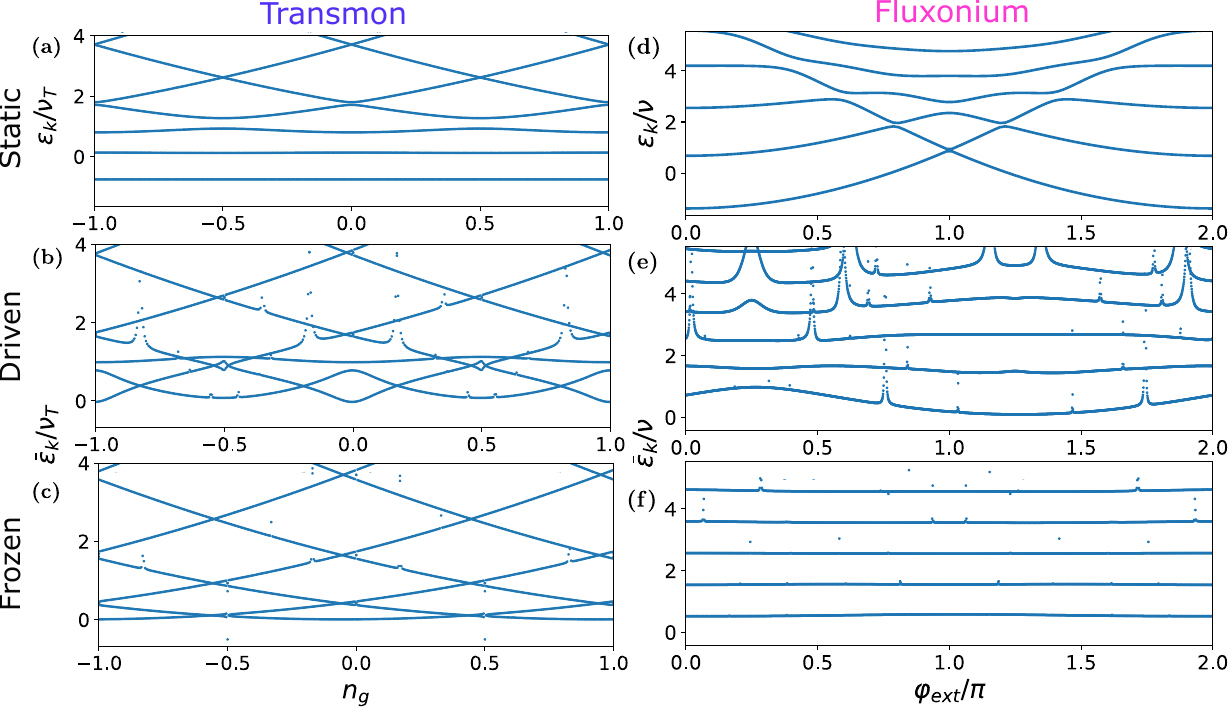}
    \caption{\textbf{Figure 5: Dependence of driven transmons and frozonium on parametric noise. } We show the eigenenergies for the {\it static} (a) transmon, and (d) fluxonium, as a function of $n_g$ and $\vp_{\tn{ext}}$, respectively and the period averaged quasi-energies, $\bar{\ve}_k$, for a {\it driven} (b)-(c) transmon, and (e)-(f) fluxonium, as a function of $n_g$ and $\vp_{\tn{ext}}$, respectively. The $\bar{\ve}_k$ are normalized by the transmon oscillator energy $\nu_T/h=\sqrt{8E_{J}E_{c}}/h \approx 3.25 \text{ GHz}$ in panels (a)-(c), and by the fluxonium oscillator energy $\nu$ (\eqref{eq:osc_freq}) in panels (d)-(f). The drive frequency is $\omega/\nu = 9.23$ for both. (b) and (e) are away from freezing, with $\alpha = 1.5$. (c) and (f) are at freezing, with $\alpha = 2$. Driven transmons at freezing appear somewhat less sensitive to variations in charge noise than transmons driven away from freezing. Nonetheless, both cases of driven transmon remain more sensitive to charge noise than static transmons. In contrast driven fluxonium away from freezing is less sensitive to variations in flux noise than static fluxonium (with the exception of resonances), and frozonium (f) is almost entirely independent to variations in flux noise.}\label{fig:E_vs_phi_ext}
\end{figure*}

We now return to one of the main practical advantages of the Floquet protocol in the vicinity of freezing, namely of an enhanced robustness to external noise. Before describing the results for the driven fluxonium due to variations in the external flux, $\vp_{\rm{ext}}$, let us briefly comment on the results for the driven transmon [\eqref{Hamiltonian1}] in the presence of variations in $n_g$. 
Static transmons are known to be resistant to fluctuations in charge noise since the large Josephson energy (relative to charging energy) exponentially suppresses the charge dispersion of the low-lying eigenstates~\cite{koch_charge-insensitive_2007}. 
On the other hand, driven transmons at generic values of the drive amplitude become dramatically more susceptible to effects of charge noise~\cite{koch_charge-insensitive_2007}. 
At freezing points, the driven transmons behave as a pure charging island, which have quadratic dependence on offset charge.
Within the Floquet-Magnus expansion, this can be understood as follows. The term in the effective Hamiltonian stemming from the external charge noise is of the form $(n_g\widehat{n})$, which commutes with the zeroth-order term $(4E_c \widehat{n}^2)$. The spectrum of the dynamically frozen transmon is then governed by the higher order corrections. To demonstrate this numerically, instead of analyzing the effects of a fully time-dependent noise $n_g(t)$, we assume that variations in charge noise occur on a slower timescale compared to the relevant energy scales of our driven system~\cite{de_leon_materials_2021,serniak_direct_2019}. 
Within this adiabatic approximation, we treat $n_g$ as a quenched variable.

In Figs.~\ref{fig:E_vs_phi_ext}a-c we compare the results for $\bar{\ve}_k$ for the (un)driven single transmon
as a function of $n_g$ at and away from freezing. The effect of variations in $n_g$ can be understood as changing the level-spacing of the computational eigenstates leading to phase-errors. As anticipated, the weak $n_g$ dependence of the low energy levels of the static transmon [panel (a)] becomes stronger when the transmon is driven, whether away from freezing [panel (b)] or at freezing [panel(c)].

Turning now to the driven fluxonium, we vary $\vp_\text{ext}$ between $[0,2\pi)$ instead of keeping it fixed at $\vp_\text{ext}=\pi$. 
As noted in the introduction, variations in $\vp_\text{ext}$ change the level-spacing of the computational eigenstates leading to phase errors. 
Experimental protocols that provide robustness against any variations in the energy difference between the computational eigenstates as a function of $\vp_\text{ext}$ are thus highly desirable. 
As in our treatment of charge noise for transmons, instead of analyzing the effects of a fully time-dependent noise, $\vp_\text{ext}(t)$, we make use of an adiabatic approximation below. 
We again assume that variations in $\vp_\text{ext}$ occur on a slower timescale compared to the relevant energy scales of our driven system,  and treat $\vp_\text{ext}$ as a quenched variable.

The external flux shifts the Josephson and inductive terms relative to each other and thus can be chosen to only enter in the Josephson term~\cite{vool_introduction_2017} (see \eqref{eq:H_mov}).
At freezing points, the Josephson nonlinearities are strongly suppressed by powers of $1/\omega$. The effect of the nonlinearities on the period averaged quasi-energies is analytically expected to be $O(1/\omega^2)$ for both the triangular-wave and sine wave drives (see Supplementary material).
Furthermore, for the case of the triangular-wave drive at the $k-$th freezing point (to second order in the Floquet-Magnus expansion), we find that the \textit{quasi-energy} gap between the computational states (modulo $\omega$), normalized by $\nu$, is analytically expected to scale with $\vp_\text{ext}$ as
\begin{align}
    \frac{\epsilon_1-\epsilon_0}\nu =\ &1-\frac{1}{k^2\omega^2}\frac{E_C}{E_{\rm{L}}}e^{-\sqrt{8E_C/E_{\rm{L}}}}E_{\rm{J}}^2\cos(2\vp_\text{ext})\label{eq:E_mag_second_order}\\
    &+ O\left(\frac{1}{\omega^3}\right)\nonumber\,,
\end{align} (see Supplementary material).
We note that the leading $\vp_\text{ext}$-dependent term in this expression is exponentially suppressed in $E_C/E_{\rm{L}}$. 
The same exponential suppression of dispersion also appears for higher energy levels that we computed (such as $\epsilon_2-\epsilon_1$), suggesting that the anharmonicity of frozonium at freezing points is also exponentially suppressed at freezing (see Supplementary material for further discussion).
We note that this differs from the behavior of transmons where the charge-noise dispersion is known to experience an exponential suppression (by $E_{\rm{J}}/E_{C}$) while the anharmonicity does not \cite{PhysRevA.76.042319}.

Putting all of this together, we thus anticipate that the phenomena of freezing will be robust against slow flux noise, and experience a lower tendency towards phase errors compared to the undriven system. 
In Figs.~\ref{fig:E_vs_phi_ext}d-f, we compare the results for $\bar{\ve}_k$ for the (un-)driven fluxonium as a function of $\vp_\text{ext}$ at and away from freezing. 
There is a clear tendency for the different quasi-energy states to become largely insensitive to variations in $\vp_\text{ext}$ at freezing [panel (f)], compared to the undriven problem [panel (d)]. 
Away from freezing [panel (e)], while there are regions of $\vp_\text{ext}$ where the quasi-energies have minimal dependence on $\vp_\text{ext}$, there are also isolated resonances, as discussed previously. 
We note that somewhat similar physics to this has been described in a past work in the context of the computational states of a flux driven fluxonium\cite{Huang_2021}. Frozonium has some crucial advantages over these previous set up. By including an additional coherent charge drive, frozonium extends this resistance to flux noise to the full bosonic spectrum as well as allowing for exponential resistance to flux noise in the case of the triangular wave drive.

To summarize, we note that a clear benefit of frozonium over driven transmons is the wider regime of protection against both flux and charge noise, in addition to the ability to tune the strength of induced nonlinearity away from freezing. Investigating this advantage in experiments in both driven transmons and frozonium in the vicinity of freezing points is an exciting future direction.

\section{Discussion}\label{sec:outlook}

In this work we have analyzed the dynamical properties of a novel frozonium circuit, consisting of an inductively shunted Josephson junction under the effect of a Floquet drive. 
We have used a combination of numerical exact diagonalization and analytical calculations based on the Floquet-Magnus expansion for revealing the properties of frozonium over a wide range of drive amplitudes and frequencies, respectively. 
At special ratios of the drive amplitude and frequency, we have identified a set of isolated freezing points where the effective quantum dynamics of the frozonium is well-approximated by a simple harmonic oscillator, supplemented by anharmonic nonlinear corrections that are suppressed in high powers of the drive frequency. 
Tuning away from these freezing points allows for additional continuous control of the degree of quantum nonlinearity. 
Treating the external flux-noise as a quenched variable in the adiabatic approximation, we have also demonstrated robust protection of the frozonium from decoherence. 
We find that the parametric dependence on magnetic flux is strongly suppressed at the freezing point, which suggests protection against dephasing from flux noise. We remark that result, using large classical flux fluctuations, connects to similar reductions in parametric flux dependence in the condition of large quantum phase fluctuations~\cite{pechenezhskiy_superconducting_2020}. Additionally, we have clarified the phenomenology of the dynamically generated isolated resonances between the quasi-energy states separated by the drive frequency, which are undesirable from the point of view of quantum control.

We see two important practical applications of the dynamical control we have developed throughout this manuscript. 
Inspired by dynamical freezing and reduction in chaos in many-transmon systems \cite{mukherjee2024}, the first is to adapt this control to protect quantum information in many coupled frozonium from the spread of chaos over long times.
At freezing points, the frozonium becomes approximately integrable. 
By tuning to these special points, a non-integrable, chaotic, coupled fluxonium system can be converted to be approximately integrable and non-chaotic. 
A possible impediment to this proposal in the case of transmon systems was due to the presence of noisy offset charge, which frozonium does not exhibit with respect to flux. 
Our theoretical analysis of the driven fluxonium (and previously the driven transmons \cite{mukherjee2024}) has treated the problem as a closed and unitary quantum mechanical system. One interesting future direction would be to include the coupling of these driven many-body systems to an external bath, and analyze the robustness of the freezing phenomenology in the presence of weak dissipation.

Finally, we remark on a second potential application, namely bosonic quantum control. 
Over recent years, bosonic quantum computing has drawn significant interest due to its ability to implement several categories of error correcting codes and accomplish quantum simulation~\cite{ma_quantum_2021,liu_hybrid_2024,bosoniccontrol1,bosoniccontrol2}.
However, universal control schemes usually require coupling an ancilla qubit to the bosonic mode. As a result, the gate speeds in such architectures are fundamentally limited by the dispersive coupling strength between the ancilla and the bosonic mode. Furthermore, during the gates, even high-Q bosonic modes become generally susceptible to errors in the low-Q ancilla~\cite{ma_quantum_2021}.

The type of dynamical control over the fluxonium circuit described here therefore presents a new direction for bosonic control in which only a single dynamical mode is required.
At freezing points we show that driven fluxonium is well-approximated by a harmonic oscillator, and nonlinearity is introduced when tuning the drive parameters away from the freezing point. 
Frozonium has the potential to circumvent the limitations of traditional bosonic architectures by dynamically tuning the mode’s intrinsic nonlinearity. 
This feature enables us to protect quantum information using the harmonic regime and tune to the nonlinear regime to perform readout or gate operations, eliminating the need for an ancilla qubit. 
While such gates may not be error-correctable at the single mode level, they can potentially be extremely fast and hence less susceptible to errors compared to traditional bosonic operations due to the large in-situ non-linearity of frozonium. 
Moreover, realizing gates in the standard bosonic architecture requires performing simultaneous drive on two circuit elements, making it a complex optimal control problem that is inherently susceptible to control errors~\cite{ma_quantum_2021,liu_hybrid_2024}. 
In contrast, frozonium uses a single nonlinear mode for gates, simplifying the control scheme. We leave a targeted exploration of the power of this kind of control to a future study.

We also note that some recent works discuss the possibility of drive-induced quasiparticle excitations as a source of noise in Floquet superconducting qubits \cite{chowdhury2025theoryquasiparticlegenerationmicrowave,kishmar2025quasiparticleinduceddecoherencedrivensuperconducting}. 
Given the results of these works, frozonium lives in rather favorable regime with respect to quasiparticle creation for several reasons. 
In particular, frozonium requires a comparatively small drive phase amplitude to reach the first freezing point, and quasiparticle generation is found to decrease sharply at lower drive phase amplitudes \cite{chowdhury2025theoryquasiparticlegenerationmicrowave}. Frozonium utilizes an inductive term to cap the extent to which the drive can affect the $\hat\varphi$-dependent potential, preventing highly-detrimental unstable-state localization \cite{chowdhury2025theoryquasiparticlegenerationmicrowave}. Finally, frozonium does not require a large Josephson energy to function, further reducing quasiparticle generation which scales with $E_J$ \cite{chowdhury2025theoryquasiparticlegenerationmicrowave,kishmar2025quasiparticleinduceddecoherencedrivensuperconducting}. For these reasons, we find it likely that quasiparticle
generation in frozonium will not be a dominant source of noise. 

It has also been pointed out that quasiparticle generation is exponentially suppressed as a function of $\omega/\Delta$ and thus will be significantly reduced by using materials with larger gaps \cite{chowdhury2025theoryquasiparticlegenerationmicrowave,kishmar2025quasiparticleinduceddecoherencedrivensuperconducting}.
Implementing superconducting qubits in materials such as niobium \cite{PhysRevApplied.21.024047} or tantalum \cite{PhysRevX.13.041005,Place_2021,bothara2025highfidelityqndreadoutmeasurement,Li_2025} is thus a promising direction to suppress quasiparticle generation in all forms of driven superconcduting circuit. We also note that a recent experimental work \cite{ann2025evidenceunexpectedlylowquasiparticle} has found a significantly lower quasiparticle generation rate than predicted in theory, which is an interesting problem for future study. Further investigation of the degree to which frozonium is resistant to quasiparticle generation and into methods to suppress quasiparticle generation for general driven superconducting circuits is an interesting and important direction for future work.

\section{Methods}\label{sec:methods}

We discuss here the methods used to perform the exact diagonalization (ED) computations throughout this manuscript. We have made use of the QuSpin PYTHON package~\cite{weinberg2017quspin,weinberg2019quspin} to perform the ED analysis. We write the Hamiltonian in terms of bosonic raising and lowering operators $b^\+ ,b$,
\begin{align}
    \widehat{\vp}&=\left(\dfrac{2E_C}{E_{\tn{L}}}\right)^{1/4} [b^\+ +b],\\
    \widehat{n}&=\dfrac{i}{2}\left(\dfrac{E_{\tn{L}}}{2E_C}\right)^{1/4}[b^{\dagger}-b],\label{eq:bosonic_operator}
\end{align} 
where $[b,b^\+ ]=1$. This permits a number basis representation $\{\ket{n}\}$ of our Hilbert space, which we truncate to keep only the lowest $d_H$ states. Our operators written in terms of $b^\+ , b$ thus become $d_H\times d_H$ dimensional matrices. We take $d_H=70$ throughout unless otherwise noted. We find that the time averaged Floquet eigenenergies converge well for this Hilbert space dimension in the presence of a drive (see Supplementary material).

To determine the matrix elements of the nonlinear $\cos(\widehat{\vp})$ term in the Hamiltonian we apply an exact matrix cosine to the truncated operator. Specifically, for a given Hermitian operator written as a matrix $\hat{M}$ we note that,
\begin{align}
    \cos{\hat{M}} = \cos(\hat{U}^{-1} \hat{\Lambda}\hat{U}) = \hat{U}^{-1}\cos(\hat{\Lambda}) \hat{U},
\end{align} 
where $\hat{U}$ is the transformation that diagonalizes the matrix, and $\hat{\Lambda}$ is the resulting diagonal matrix consisting of the eigenvalues of $\hat{M}$. The diagonal matrix $\cos(\hat{\Lambda})$ can then be calculated via the element-wise application of cosine. 

By calculating the exact matrix cosines only after truncating the operator, we keep the important periodic structure of the cosine. 
In contrast, this periodic structure is broken if we were to instead use the exact cosine matrix elements $\bra{m}\cos\hat\varphi\ket{n}$ computed using the full Hilbert space (as determined in Ref.\cite{Catelani_2011}) and only then truncate the resulting operator. 
We find that by breaking periodicity, this latter method gives rise to additional fictitious eigenmodes that live in the middle of the fluxonium spectrum and depend strongly on $d_H$.
By performing the correct order of limits, the method we make use of entirely avoids this.

To compute the time evolution of our system up to late times, we make use of the Floquet theorem. Specifically, in the presence of a time periodic Hamiltonian $\mathcal{H}(t)$ with frequency $\omega$ and period $T = 2\pi/\omega$, the time dependent eigenvectors of $\mathcal{H}(t)$ can be written as $\ket{u_k(t)}=e^{-i\ve_k t}\ket{\phi_k(t)}$ for Floquet eigenvector $\ket{\phi_k(t)}$ satisfying $\ket{\phi_k(t+T)} = \ket{\phi_k(t)}$, where the quasi-energy $\ve_k$ is defined up to modulo $\omega$. The Floquet eigenvectors $\{\phi_k(0)\}$ can be calculated via diagonalizing the time ordered evolution operator over one period given by,
\begin{align}
    U(T,0) = \left(\mathcal{T}\exp\left(-\int_0^{T}dt'i{\cal{H}}(t')\right)\right).
\end{align} 
We can then determine the stroboscopic evolution of a state $\ket{\psi}$ by projecting onto the Floquet eigenvector basis and evolving,
\begin{align}
    \ket{\psi(t = n T)} = \sum_k e^{-i\ve_k nT}\braket{\phi_k(0)|\psi(0)}\ket{\phi_k(0)},
\end{align} 
where $n\in\mathbb{Z}$. By using this method, we are able to make the computational time needed to determine $\ket{\psi(t = n T)}$ fully independent of the value of $nT$. Finally, to compute the period-averaged quasienergies (\eqref{eq:Period_avg_E}), we discretize the integral into $50$ equally spaced time slices, and average over the integrand computed at each of these slices.

For the numerical simulations of the transmon Hamiltonian, we work in the charge basis~\cite{QiskitMetalTutorial}, where the full static part of the Hamiltonian is expressed as,  
\begin{equation*}  
    H_{\text{static}}=4E_{c}(\hat{n}-n_{g})^2-\frac{E_{J}}{2}\sum_{n}\bigg(\ket{n}\bra{n+1}+ \tn{h.c.} \bigg),  
\end{equation*}  
with $\hat{n}=\sum_{n=-\infty}^{\infty}\ket{n}\bra{n}$. The Floquet drive is given by $f(t)H_{\rm{drive};\widehat{n}}$, where $f(t)$ is given by \eqref{eq:sq_transmon}. For numerical simulations, we apply a cutoff at $n=10$ thus keeping $21$ charge states. We have verified that increasing the cutoff has a negligible effect on the low-lying excitation spectrum in the regime of interest.

\section*{Resource Availability}

\subsection{Lead contact} Further information and requests should be directed to and will be fulfilled by Debanjan Chowdhury (debanjanchowdhury@cornell.edu).

\subsection{Materials availability} This study did not generate new materials.

\subsection{Data and Code Availability}

\begin{itemize}
    \item All the theoretical data generated in this study have been deposited at Zenodo and are publicly available as of the date of publication: \url{https://zenodo.org/records/18394019}
    \item The custom computer codes used to generate the results reported in this paper are available from the lead contact upon request.
\end{itemize}

\section*{Acknowledgments}
We thank Michel Devoret for insightful discussions. KL acknowledges that this material is based upon work supported by the National Science Foundation Graduate Research Fellowship under Grant No. DGE – 2139899. HG is supported by a Wilkins postdoctoral fellowship at Cornell University. DC, RM and HG are supported in part by a New Frontier Grant awarded by the College of Arts and Sciences at Cornell University, and by a Sloan research fellowship to DC from the Alfred P. Sloan foundation.

\section*{Author Contributions}
V.~F. and D.~C. determined the motivations for adding the inductive shunt; K.~L., R.~M., and H. G. performed the theoretical computations under supervision of D.~C. with periodic input from S.~R. and V.~F.; K.~L. performed the frozonium numerical simulations, and R.~M. performed the transmon numerical simulations; K.~L. and D.~C. wrote the manuscript with input from all authors.\\

\section*{Declarations of Interest}
The authors declare no competing interests

\appendix

\section{Additional Details: Floquet-Magnus Expansion}\label{app:Magnus}
We supplement the results of Sec.~\ref{sec:FM} with additional details here, for both the triangular and sine-wave drives, respectively. Let us note at the outset that the Floquet-Magnus expansion is a high-frequency perturbative expansion that equates a periodically driven \textit{time-dependent} Hamiltonian into a \textit{time-independent} one over one Floquet cycle and that the resulting series is not formally convergent. For quantum problems with an unbounded Hilbert space, the formalism is not strictly justified, but nevertheless provides a useful reference point for describing the exact numerical results. We also reiterate that the Floquet quasi-energies are only defined modulo $\omega$, leading to an energy ``band-folding" for quasi-energies that exceed the drive frequency. The resonances between the different Floquet blocks (as discussed e.g. in Fig.~\ref{fig:mag_resonance}) represent important quantum mechanical processes, that also controls the non-trivial thermalization dynamics of driven problems \cite{PhysRevLett.132.100401}.

Returning now to the discussion in Sec.~\ref{sec:FM}, let us compute the first two terms in the Floquet-Magnus expansion,
\begin{subequations}\label{eq:Magnus_exp}
\begin{align}
    \mathcal{H}_{\tn{eff}}^{(0)} =\ &\frac{1}{T}\int_0^T dt ~{\cal{H}}_{\tn{mov}}(t),\\
   \mathcal{H}_{\tn{eff}}^{(1)} =\ &\frac{1}{2iT}\int_0^Tdt_1\int_0^{t_1}dt_2 ~[{\cal{H}}_{\tn{mov}}(t_1), {\cal{H}}_{\tn{mov}}(t_2)],\label{eq:Hmag1}
\end{align}
\end{subequations} 
where ${\cal{H}}={\cal{H}}_{\text{mov}}$ (\eqref{eq:H_mov}). Let $F_m$ be the Fourier coefficients of $e^{i\Theta(t)}$,
\begin{align}
    e^{i\Theta(t)} = \sum_m F_m e^{im\omega t}\label{eq:fourier_def}.
\end{align} With this notation,
\begin{subequations}
\begin{align}
    {\cal{H}}_{\text{eff}}^{(0)} &= 4E_c\hat n^2 + \frac{E_{\tn{L}}}{2} \hat\vp^2 - \frac{E_{\tn{J}}}{T}\int_0^Tdt\cos(\hat\vp-\Theta(t))\\
    &= 4E_c\hat n^2 + \frac{E_{\tn{L}}}{2} \hat\vp^2 -\frac{E_{\tn{J}}}{2}\left(F_0^*e^{i\hat\vp} + F_0e^{-i\hat\vp}\right).\label{eq:H0_mag}
\end{align}  
\end{subequations}
To calculate ${\cal{H}}_{\text{eff}}^{(1)}$, we must integrate commutators of ${\cal{H}}_{\text{mov}}$ at different times. The only time dependent component of ${\cal{H}}_{\text{mov}}$ is the transformed Josephson energy term. Furthermore,
\begin{align}
    \left[\frac{E_{\tn{L}}}{2}\hat\vp^2, E_{\tn{J}}\cos(\hat\vp-\Theta(t))\right] = 0.
\end{align} 
The commutators that appear in Eq.~ \eqref{eq:Magnus_exp} are thus independent of $E_{\tn{L}}$. As a result, the first order term of the Magnus expansion for our fluxonium system are identical to the equivalent terms in the driven transmon system \cite{mukherjee2024}, leading to
\begin{subequations}
\begin{align}
    {\cal{H}}_{\text{eff}}^{(1)} &= \frac{2}{\omega} E_cE_{\tn{J}}\left[B(2\hat n+1)e^{-i\hat\vp} + \text{h.c.}\right]\\
    &=\frac{2}{\omega} E_cE_{\tn{J}}\left[\hat n(Be^{-i\hat\vp} + B^*e^{i\hat\vp}) + \text{h.c.}\right],
\end{align}
\end{subequations}
where
\begin{align}
    B = \sum_{m\ne 0} \frac{F_m}{m}.
\end{align} 
This expression gives rise to the $O(1/\omega)$ suppression of Magnus corrections mentioned in the main text. 

A notable formal consideration is what \textit{dimensionless} constant these corrections are small in. While a precise expression is complicated due to form of the Magnus expansion, we can provide a conservative estimate as follows. The $k^\text{th}$ order term of the Magnus expansion contains $k$ order commutators between $H$ and itself at different times. This will give rise to monomials in terms of $E_C,E_L,E_J$ or order $k+1$. The $k^\text{th}$ order term of the Magnus expansion is also suppressed by $\omega^k$. As the correction should have dimensions of energy, we must have 
$\omega^k$ be large relative to the $k^\text{th}$ order monomials in terms of $E_C,E_L,E_J$. Conservatively, it is thus sufficient for $\omega \gg \text{max}\{E_C, E_L, E_J\}$. For the parameter regime used throughout this work, an appropriate dimensionless constant is thus $\omega/E_J$.

To simplify the expression above further, we need the explicit forms for $F_m$, which depend on the drive protocol. 

\subsection{Triangular wave drive} 
Let us consider the form of the drive introduced in Eq.~ \eqref{eq:tw_drive}. The Fourier coefficients defined in \eqref{eq:fourier_def} are given by

\begin{align} F_m = - \frac{i\alpha\left(-1 + (-1)^m e^{i\pi\alpha}\right)}{\pi(\alpha^2 - m^2)}.
\end{align} Note that because $F_m/m$ is an odd function of $m$, we have that $B = 0$. As a result, ${\cal{H}}^{(1)}_{\text{eff}} = 0$. Thus, from \eqref{eq:H0_mag}, at the leading order the effective Hamiltonian is given by
\begin{align}
    {\cal{H}}_{\text{eff}} =\ &4E_c\hat n^2 + \frac{E_{\tn{L}}}{2}\hat\vp^2\\
    &-\left(\frac{2}{\pi\alpha}\sin\left(\frac{\pi\alpha}{2}\right)\right)E_{\tn{J}}\cos\left(\hat\vp-\frac{\pi\alpha}{2}\right)+O\left(\frac{1}{\omega^2}\right)\nonumber.
\end{align} 
As noted previously in the main text, the leading nonlinear correction involving the $\cos\hat\vp$ vanishes at particular values of $\alpha=2n$ for $k\in\mathbb{Z}$ (\eqref{eq:tw_freezing_points}).

\begin{figure*}
    \centering
    \begin{subfigure}{0.5\textwidth}
        \centering
        \includegraphics[width=1.0\linewidth]{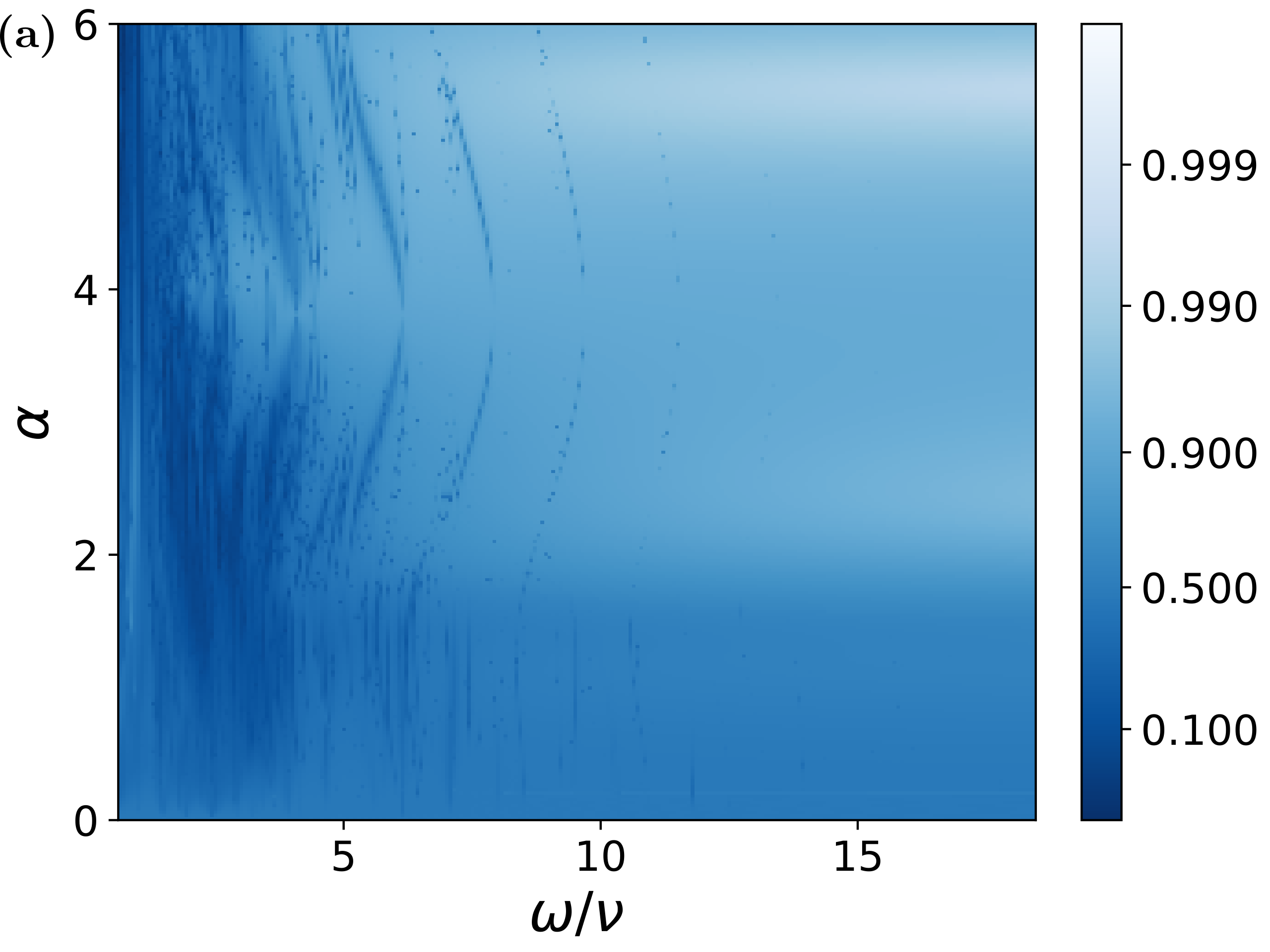}
    \end{subfigure}%
    \begin{subfigure}{0.5\textwidth}
        \centering
        \includegraphics[width=1.0\linewidth]{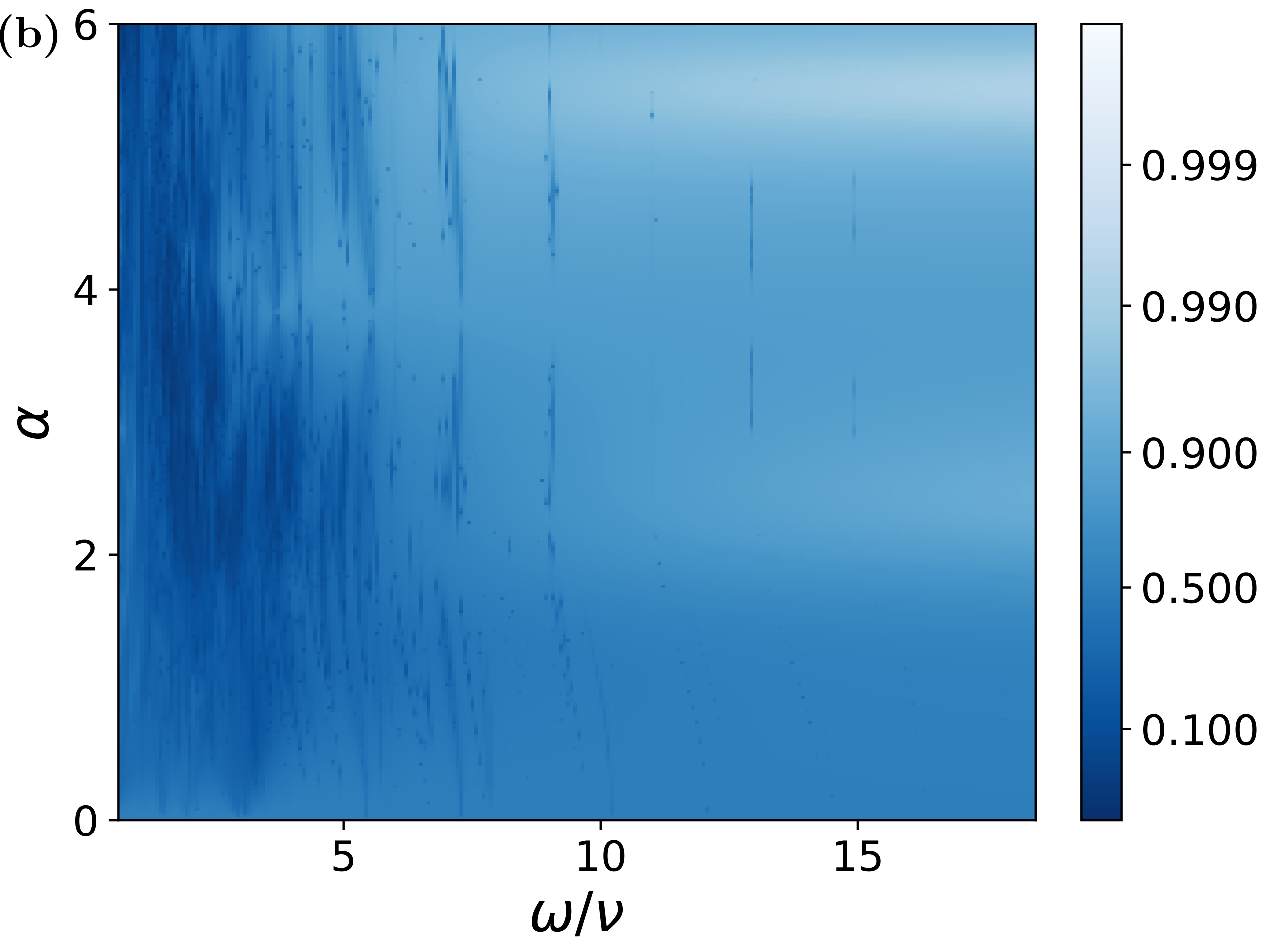}
    \end{subfigure}
    \begin{subfigure}{0.5\textwidth}
        \centering
        \includegraphics[width=1.0\linewidth]{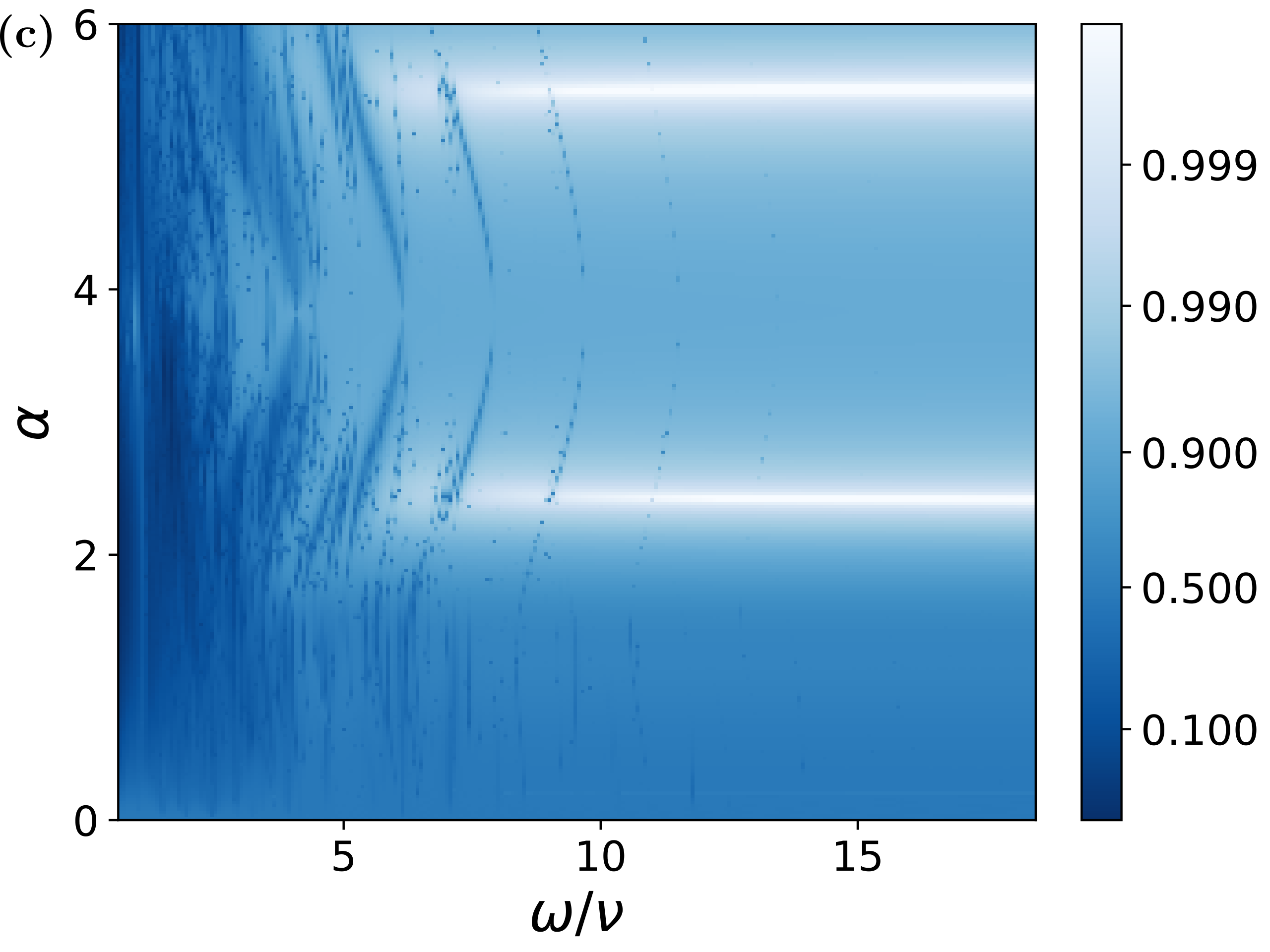}
    \end{subfigure}%
    \begin{subfigure}{0.5\textwidth}
        \centering
        \includegraphics[width=1.0\linewidth]{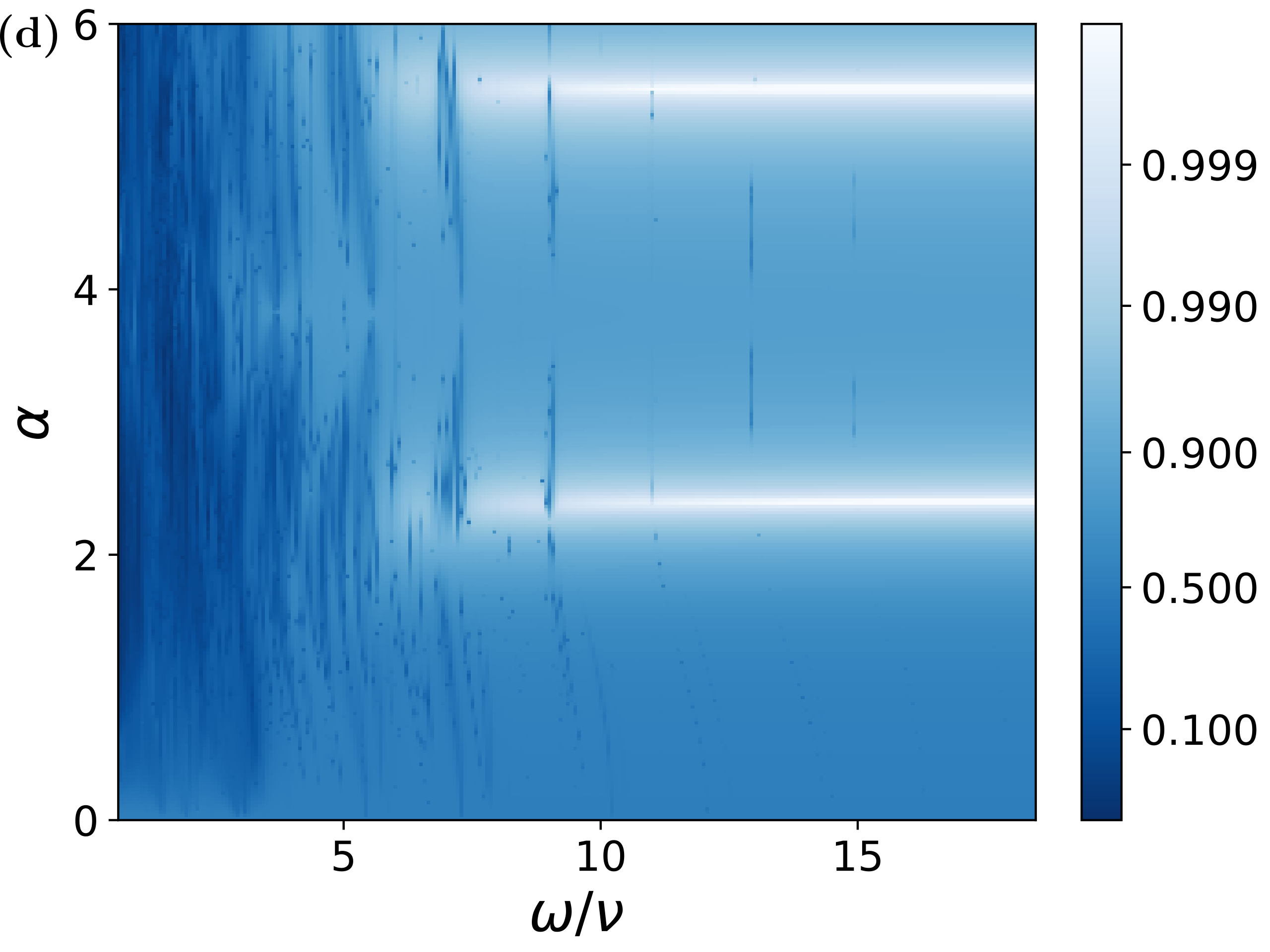}
    \end{subfigure}%
\caption{\textbf{Freezing points of sine-driven frozonium.} The IPR (Eq.~\ref{eq:IPR}) as a function of $\alpha$ and $\omega/\nu$ for a fluxonium driven via $\Theta_{\tn{sin}}(t)$ (\eqref{eq:sin_drive}) with $\vp_{\tn{ext}}=\pi$ time-averaged over $N_F=10,000$  cycles. $\ket{\Psi(0)}$ is initialized to the (a) ground state, and (b) first excited state of $H_{\tn{quad}}$ in \eqref{eq:quad}. The horizontal bands of relatively high $\text{IPR}>0.9$ centered near $\alpha = 2.4048, 5.5201$ are associated with the freezing region (see \eqref{eq:H_eff_cos}), but are  weaker than the freezing regions for the triangular wave drive (see Fig.~\ref{fig:IPR}). We use the same arctanh color scheme introduced in Fig.~\ref{fig:IPR}. The $\text{IPR}$ is shown for ${\ket{\Psi(0)}}$ initialized to the (c) ground state, and (d) first excited state of ${\cal{H}}_{\text{eff}}|_{\alpha=\alpha_k^0}$ in \eqref{eq:sin_first_order} (which includes the $O(1/\omega)$ Floquet-Magnus correction). The value of IPR in the vicinity of the same freezing points increases significantly. All the Hamiltonian parameters  are the same as in Fig. \ref{fig:IPR}.}\label{fig:IPR_cos}
\end{figure*}

\subsection{Sine wave drive} 

For the sine wave drive in \eqref{eq:sin_drive}, 
\begin{subequations}
\begin{align}
    &F_m = J_m(\alpha),~\tn{and}\\
    &B = \frac{\pi}{2}\textbf{H}_0(\alpha),
\end{align} 
\end{subequations}
where $J_m$ are Bessel functions, and $\textbf{H}_0$ is the Struve function. The effective Hamiltonian then becomes,
\begin{align}
    {\cal{H}}_{\text{eff}} =\ &4E_c\hat n^2 + \frac{E_{\tn{L}}}{2} \hat\vp^2 -J_0(\alpha) E_{\tn{J}}\cos\hat\vp\label{eq:H_eff_cos} \\
    &+ \frac{2\pi}{\omega}\textbf{H}_0(\alpha)E_c E_{\tn{J}}\left(\hat n\cos\hat\vp + \text{h.c.}\right) + O\left(\frac{1}{\omega^2}\right).\nonumber
\end{align} 
Note that at the zeros of $J_0(\alpha)$, the zeroth-order contribution from the Josephson term vanishes, leading to freezing. The nonlinear terms are suppressed in $O(1/\omega)$. The residual effective Hamiltonian at $\alpha=\alpha_k^0$, the $k^{\text{th}}$ zero of $J_0(\alpha)$, is given by
\begin{align}
    {\cal{H}}_{\text{eff}}\bigg|_{\alpha=\alpha_k^0} =\ &4E_c\hat n^2 + \frac{E_{\tn{L}}}{2} \hat\vp^2\label{eq:sin_first_order} \\
    &+ \frac{2\pi}{\omega}\textbf{H}_0(\alpha_k^0)E_cE_{\tn{J}}\left(\hat n\cos\hat\vp + \text{h.c.}\right)\nonumber\\
    &+ O\left(\frac{1}{\omega^2}\right)\nonumber
\end{align}  

As was discussed in the context of driven transmons \cite{mukherjee2024}, and will be shown in Appendix~\ref{app:sin_drive}, the $O(1/\omega)$ correction weakens the quality of freezing, but has little effect on the level spacing of the driven system.

\section{Exact Diagonalization Results for Sine Drive}\label{app:sin_drive}

We now turn to analyzing the behavior of a single fluxonium driven with a sine wave drive. We will proceed in analogy to the analysis of the triangular wave drive in the main text, and observe the impact of a non-zero first-order in $1/\omega$ term (as discussed in \eqref{eq:sin_first_order}) on the resulting behavior near freezing. As before, we set $d_H=70$, and choose $\vp_{\rm{ext}}=\pi$ unless otherwise noted.

We start by studying quantitatively the extent to which the eigenstates for fluxonium driven by a sine drive overlap with the eigenstates of $H_{\rm{quad}}$ in \eqref{eq:quad}, and the oscillator described by ${\cal{H}}_{\text{eff}}|_{\alpha=\alpha_k^0}$ described in \eqref{eq:sin_first_order}. For the former, we choose an eigenstate of $H_{\rm{quad}}$, and for the latter we similarly choose an eigenstate of ${\cal{H}}_{\text{eff}}|_{\alpha=\alpha_k^0}$. In both cases, we time-evolve the initial state with the full driven Hamiltonian, ${\cal{H}}_{\rm{mov}}(t)$ in \eqref{eq:H_mov}, and compute the $\text{IPR}$ at the end of $N_F=10,000$ cycles as in \eqref{eq:IPR}. Importantly, note that the overlap of $\ket{\Psi(t)}$ is to be evaluated with respect to $\ket{\Phi_i}$, which represents a complete eigenbasis associated with either $H_{\tn{quad}}$ in \eqref{eq:quad} , or ${\cal{H}}_{\text{eff}}|_{\alpha=\alpha_k^0}$ in \eqref{eq:sin_first_order}.

In Fig.~\ref{fig:IPR_cos}(a)-(b) we show the averaged $\text{IPR}$ as a function of $\alpha$ and $\omega/\nu$, with $\ket{\Psi(0)}$ initialized to the ground and first excited states of $H_{\rm{quad}}$, respectively. As expected, the bands of a large $\text{IPR}>0.9$ centered around the first two zeroes of $J_0$ Bessel function, $\alpha = 2.4048, 5.5201$, are clearly visible. However, unlike for the triangle-wave drive, we find a lower value of the IPR near freezing (except for large $\omega$). We conjecture that this is due to the non-vanishing first order correction in the Floquet-Magnus expansion for the sine-wave drive, which vanishes for the triangle-wave drive.

To test this hypothesis, in Fig.~\ref{fig:IPR_cos}(c)-(d), we repeat the same analysis as above but for the the ground and first excited states of ${\cal{H}}_{\text{eff}}|_{\alpha=\alpha_k^0}$ defined in \eqref{eq:sin_first_order}. Indeed, we find that the $\text{IPR}\rightarrow1$ is now significantly larger than before in the vicinity of the freezing points. Including the first order Floquet-Magnus correction term  accounts for a better agreement with the effective dynamics of the driven problem.

\begin{figure}
    \centering
    \includegraphics[width=\linewidth]{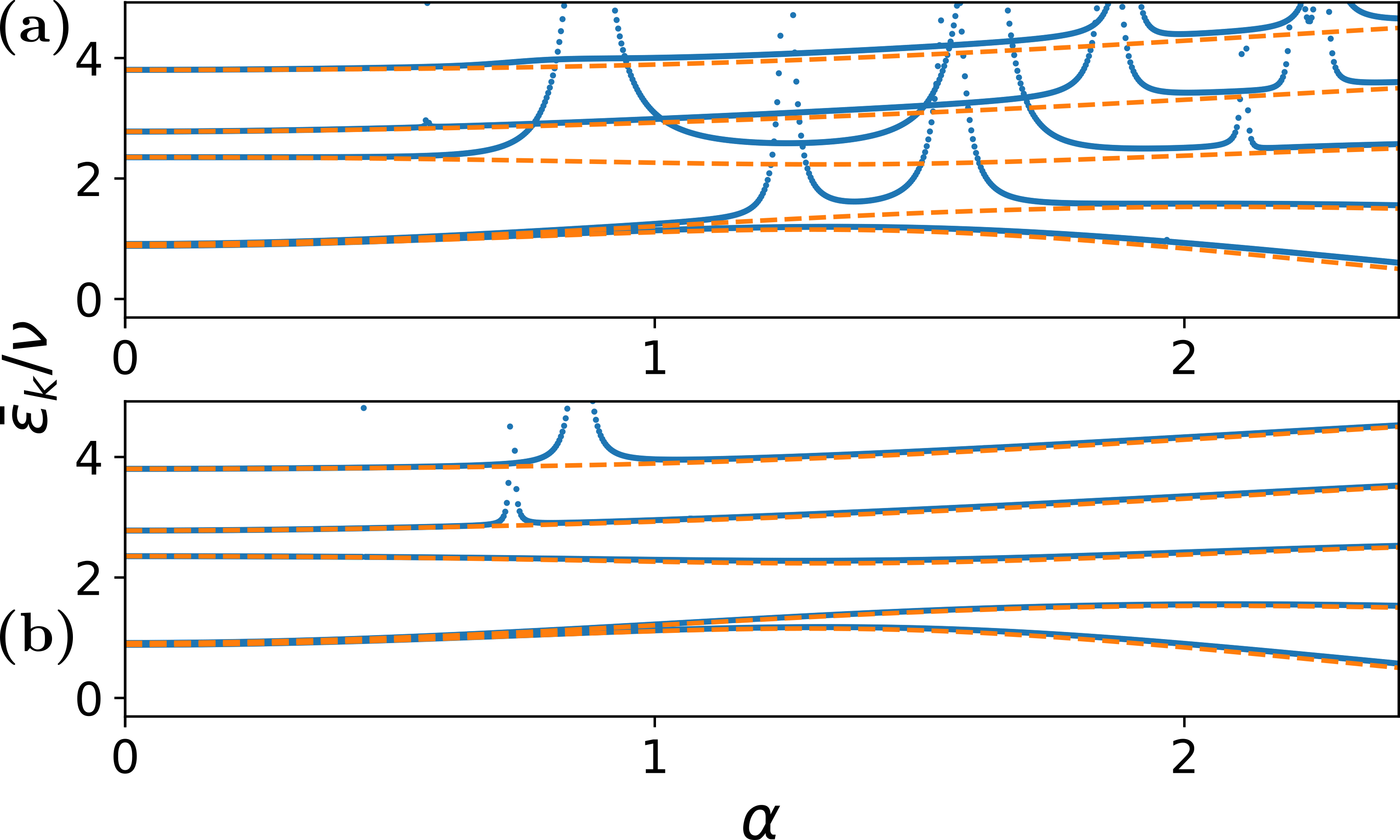}
    \caption{\textbf{Dependence of quasi-energies on drive parameters in sine-driven frozonium }The period averaged quasi-energies, $\bar{\ve}_k$, (\eqref{eq:Period_avg_E}) as a function of $\alpha$ for a driven fluxonium at (a) $\omega/\nu = 9.23$, (b) $\omega/\nu=12.31$. The result is normalized by the fluxonium oscillator energy $\nu$ (\eqref{eq:osc_freq}). All other parameters are the same as in Fig.~\ref{fig:IPR}. The right most edge of both plots corresponds to the first freezing point ($\alpha = 2.4048$). Blue data points are obtained from numerical exact diagonalization. Orange dashed lines are obtained using the zeroth-order Floquet Magnus expansion (\eqref{eq:mag_zero_sin}). 
}\label{fig:E_vs_alpha_cos}
\end{figure}

While the existence of the first order term clearly has a measurable effect on the eigenstates themselves, it has no effect on the eigenenergies of the driven system. This is because the first-order Magnus correction \eqref{eq:Hmag1} is explicitly gauge-dependent, i.e. it changes after shifting the time variable $t\to t+\Delta t$. According to the Floquet theorem, the Floquet quasienergy is independent of the gauge choice even though the eigenstates are gauge-dependent. In the following, we provide an alternative derivation using first-order perturbation theory.
 Let $\ket{\phi_k}$ be the eigenstates of the zeroth order Magnus Hamiltonian ${\cal{H}}_\text{eff}$ defined in \eqref{eq:mag_zero_sin}. The effect of the perturbative, $O(1/\omega)$ Floquet-Magnus correction on the quasi-energy of these eigenstates is given by,
 
\begin{subequations}
 \begin{align}
    E_k^{(1)} &= \bra{\phi_k}{\cal{H}}^{(1)}_\text{eff}\ket{\phi_k},\\
    {\cal{H}}^{(1)}_\text{eff} &= \frac{2\pi}{\omega}\textbf{H}_0(\alpha)E_c E_{\tn{J}}\left(\hat n\cos\hat\vp + \text{h.c.}\right).
 \end{align}
\end{subequations}

 To analyze this term let us introduce the phase parity operator $\Pi$ which acts on a wave function written in the phase representation as
\begin{align}
    \Pi\ket{\psi(\vp)} = \ket{\psi(-\vp)}\,.
\end{align} In analogy to the parity operator for the standard position and momentum representation, 
\begin{align}
    \{\Pi, \widehat{\vp}\} = \{\Pi, \widehat{n}\} = 0.
\end{align} Note that the Hamiltonian ${\cal{H}}_\text{eff}$ is even in $\widehat{n},\widehat{\vp}$, so from our anticommutation relations $[\Pi, {\cal{H}}_\text{eff}]=0$. This means that ${\cal{H}}_\text{eff}$ and $\Pi$ admit simultaneous eigenstates, so $\ket{\phi_k(\vp)}$ are all even or odd functions when written in the phase space representation.

Next note that ${\cal{H}}^{(1)}_\text{eff}$ is an odd function of $\widehat{n}$, but an even function of $\widehat{\vp}$, from our anticommutation relations $\{\Pi, {\cal{H}}^{(1)}_\text{eff}\} = 0$. As a result, acting ${\cal{H}}^{(1)}_\text{eff}$ on an eigenstate of $\Pi$ (and even or odd wavefunction) flips the parity of that state. Hence, $\bra{\phi_k}$ and ${\cal{H}}^{(1)}_\text{eff}\ket{\phi_k}$ have opposite parities and thus have zero overlap. As such, $E_k^{(1)} = 0$. The first-order term of the Magnus expansion thus only contributes at $O(1/\omega^2)$, so the eigenenergies of the system driven by the sine drive should be equally well predicted by \textit{only} the zeroth-order Magnus term as in the case of the triangle wave drive. This is in contrast to the \textit{eigenstates} of the system which are renormalized at first order perturbation theory for the sine drive versus second order for the triangular wave drive.

To see this numerically, in Fig.~\ref{fig:E_vs_alpha_cos} we plot the lowest $5$ period averaged quasi-energies (defined in \eqref{eq:Period_avg_E}) for a sine wave driven fluxonium, as a function of $\alpha$ for two different drive frequencies, $\omega$. We also compute directly the same quantity based on the zeroth- order Floquet-Magnus expansion results (\eqref{eq:mag_zero_sin}). We see that outside of resonance points, the zeroth order Magnus expansion predicts the energies of the states very well (similarly to the corresponding diagram for the triangle-wave drive, Fig \ref{fig:E_vs_alpha}).

\begin{figure}
    \centering
    \includegraphics[width=\linewidth]{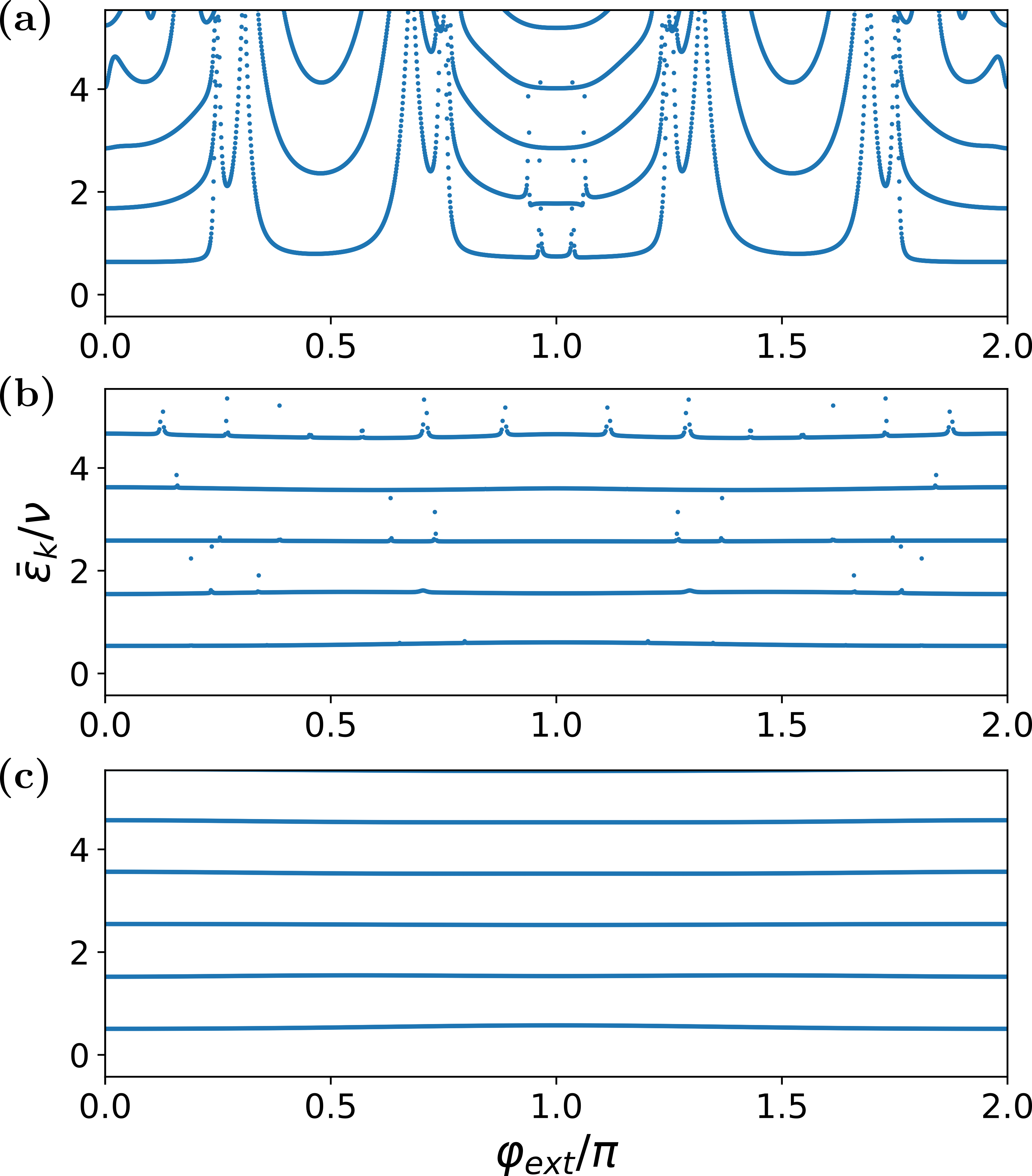}
    \caption{\textbf{Resistance of sine-driven frozonium to flux noise.} We show the period-averaged eigenenergies as a function of $\vp_{ext}$ for a single fluxonium driven by sine drive at the $\alpha=2.4048$ freezing point for (a)  $\omega/\nu = 6.15$, (b) $\omega/\nu = 9.23$, and (c) $\omega/\nu=12.31$. The result is normalized by the fluxonium oscillator energy $\nu$ (\eqref{eq:osc_freq}). All other parameters are the same as in Fig \eqref{fig:IPR}. With the exception of certain resonance points for low drive frequencies, the computational states of sine-wave driven fluxonium at freezing are almost entirely independent of external flux.  
    }\label{fig:E_vs_phi_ext_cos}
\end{figure}

Another consequence of this is that at freezing points, the period-averaged quasi-energies should be equivalent to that of $H_\text{quad}$ up to $O(1/\omega^2)$. Thus, the driven Fluxonium with a sine-wave drive should be equally robust to flux noise as the triangle-wave drive. To analyze this numerically, in Fig.~\ref{fig:E_vs_phi_ext_cos} we plot the period averaged quasi-energy of a single fluxonium qubit as a function of $\vp_\text{ext}$ at freezing for several values of $\omega$. Indeed, with the exception of the resonance points (whose number decreases with increasing $\omega$), the low-lying computational states at freezing disperse minimally with the external flux.

\section{Dependence of IPR on Drive Cycle Number}

\begin{figure}
    \centering
    \includegraphics[width=\linewidth]{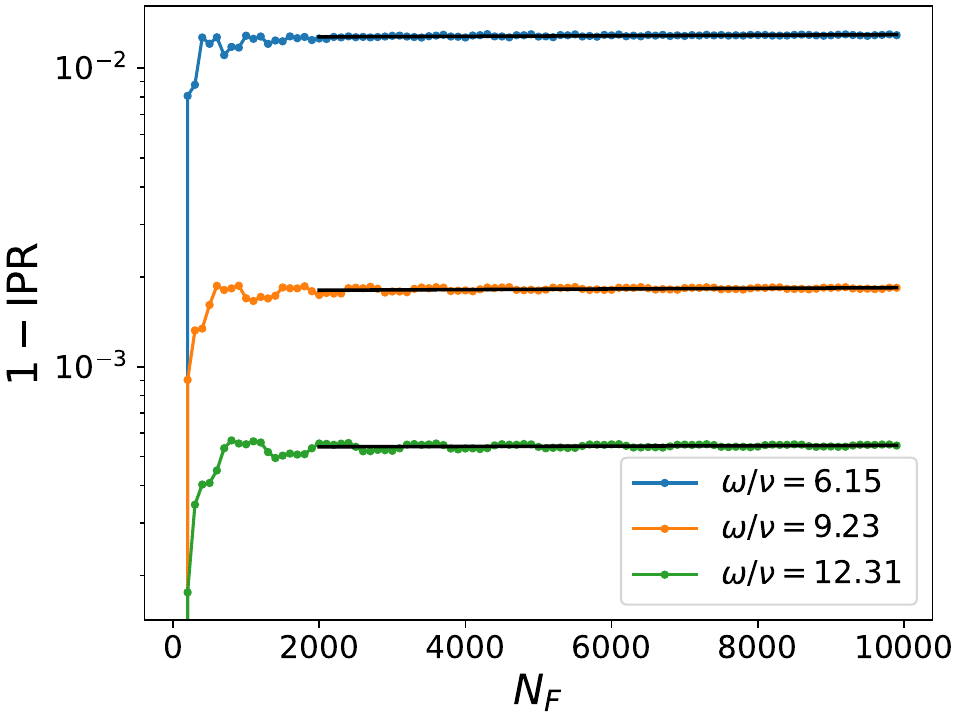}
    \caption{\textbf{Dependence of IPR on cycle number.} The averaged $1-\text{IPR}$ as a function of $N_F$ for three different values of $\omega$ for the triangular-wave drive. The system is initialized in the $\ket{0}$ state. After a short initial period of equilibration during which the IPR decays rapidly, the IPR is well described by an extremely slow exponential decay (black lines show an exponential fit).
}\label{fig:IPR_vs_NF}
\end{figure}

In Fig.~\ref{fig:IPR} and Fig.~\ref{fig:IPR_cos}, we show $1-\text{IPR}$ averaged over $N_F = 10000$ drive cycles. However, we also expect the IPR itself to be dependent on $N_F$. We now briefly discuss this dependence. In Fig.~\ref{fig:IPR_vs_NF} we plot the averaged IPR (defined in \eqref{eq:IPR}) as a function of $N_F$ for the triangular-wave drive. The system is initialized in the $\ket{0}$ state and we show results for $\omega/\nu = 6.15, 9.23, 12.31$.

We find that the system undergoes a short period of equilibration during which time the IPR decreases rapidly. The extent of this decrease is controlled by the magnitude of $\omega$ with larger frequencies giving rise to smaller drops in the IPR. After this, the IPR is well described by an extremely slow exponential decay (see black lines for such a fit).

\section{External-Flux Dependence}\label{app:dispersion}

As discussed in Sec.~\ref{sec:noise} and Appendix~\ref{app:sin_drive}, at freezing points the $\vp_\text{ext}$ dependent terms in the effective Hamiltonian are suppressed by a factor of $1/\omega^2$. Furthermore, we have observed that this leads to a qualitative reduction in the effect of flux dispersion on the period-averaged quasi-energies of the system. Nonetheless, the form of the $\vp_\text{ext}$ dependent terms in the effective Hamiltonian of the frozen system differs significantly from those in the undriven system -- potentially leading to differences in the form of the system's dependence on $\vp_\text{ext}$. In this appendix we discuss the analytic scaling of the quasienergies for the computational states of frozen system on $\varphi_\text{ext}$. For simplicity, we focus on the case when the system is driven by the triangular wave drive.

\begin{figure}
    \centering
    \includegraphics[width=\linewidth]{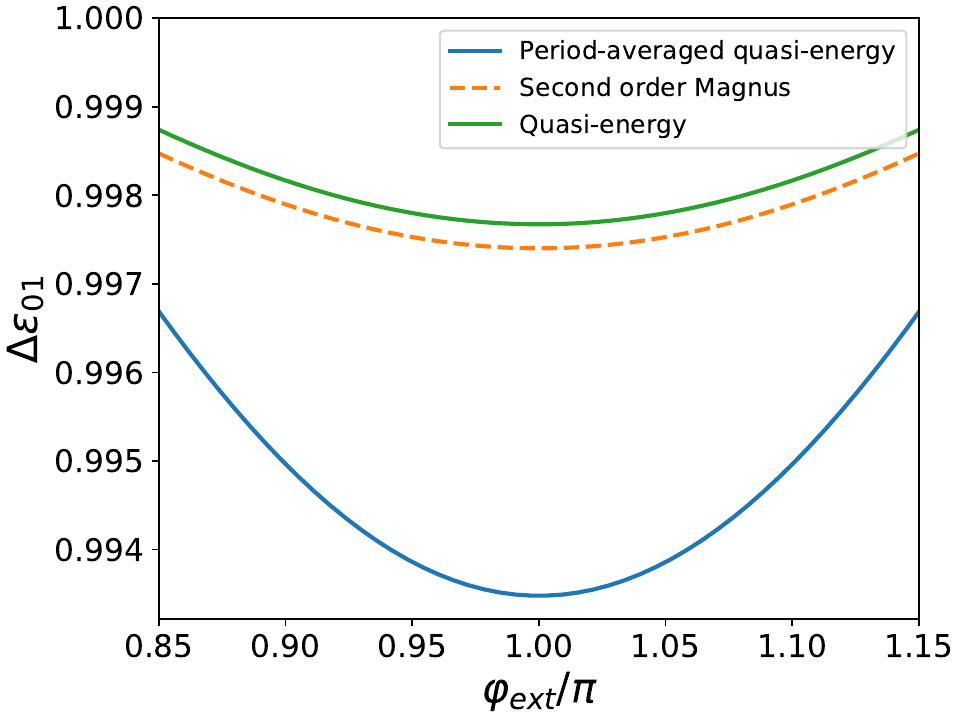}
    \caption{\textbf{Comparison of Quasi-energy with Second Order Magnus.} The normalized energy gap, $\Delta\epsilon_{01}$, between the computational states (see \eqref{eq:E_gap}) as a function of $\vp_\text{ext}$ near $\pi$-flux for frozonium driven by the triangular wave drive at the $\alpha=2$ freezing point. The period-averaged quasi-energy gap (blue), the analytical prediction (orange) of the Floquet-Magnus expansion to $O(1/\omega^2)$, and the quasi-energy gap (not period-averaged) up to modulo $\omega/\nu$ (green). The period-averaged vs. non-averaged quasi-energy gaps differ at $O(1/\omega^2)$ \cite{mukherjee2024instanton}. We choose $\omega/\nu=9.23$, and all other parameters are as in Fig.~\ref{fig:IPR}. 
}\label{fig:dispersion}
\end{figure}

\begin{figure*}[t]
    \centering
    \begin{subfigure}{0.5\linewidth}
        \centering
        \includegraphics[width=1.0\linewidth]{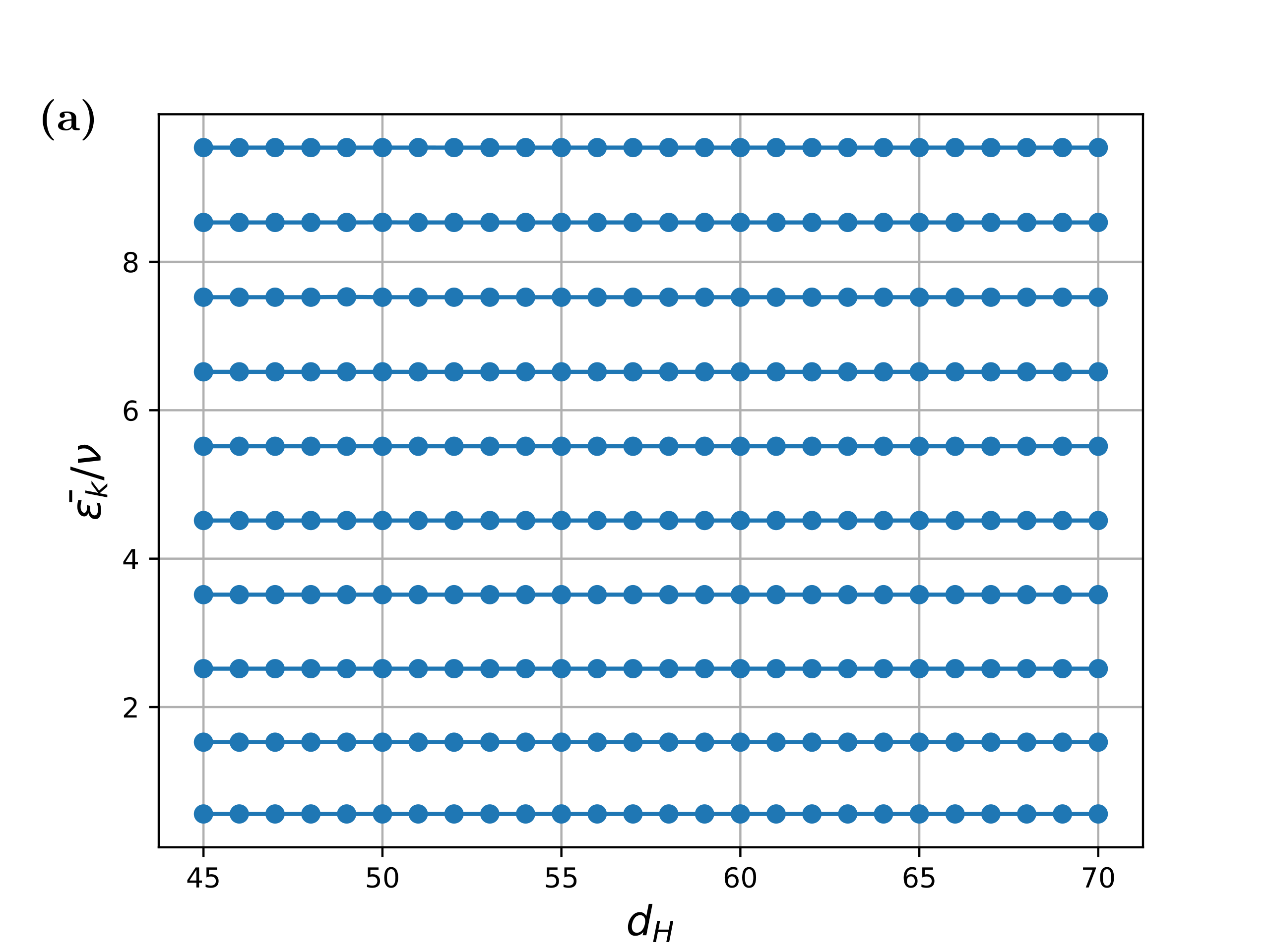}
    \end{subfigure}%
    \begin{subfigure}{0.5\linewidth}
        \centering
        \includegraphics[width=1.0\linewidth]{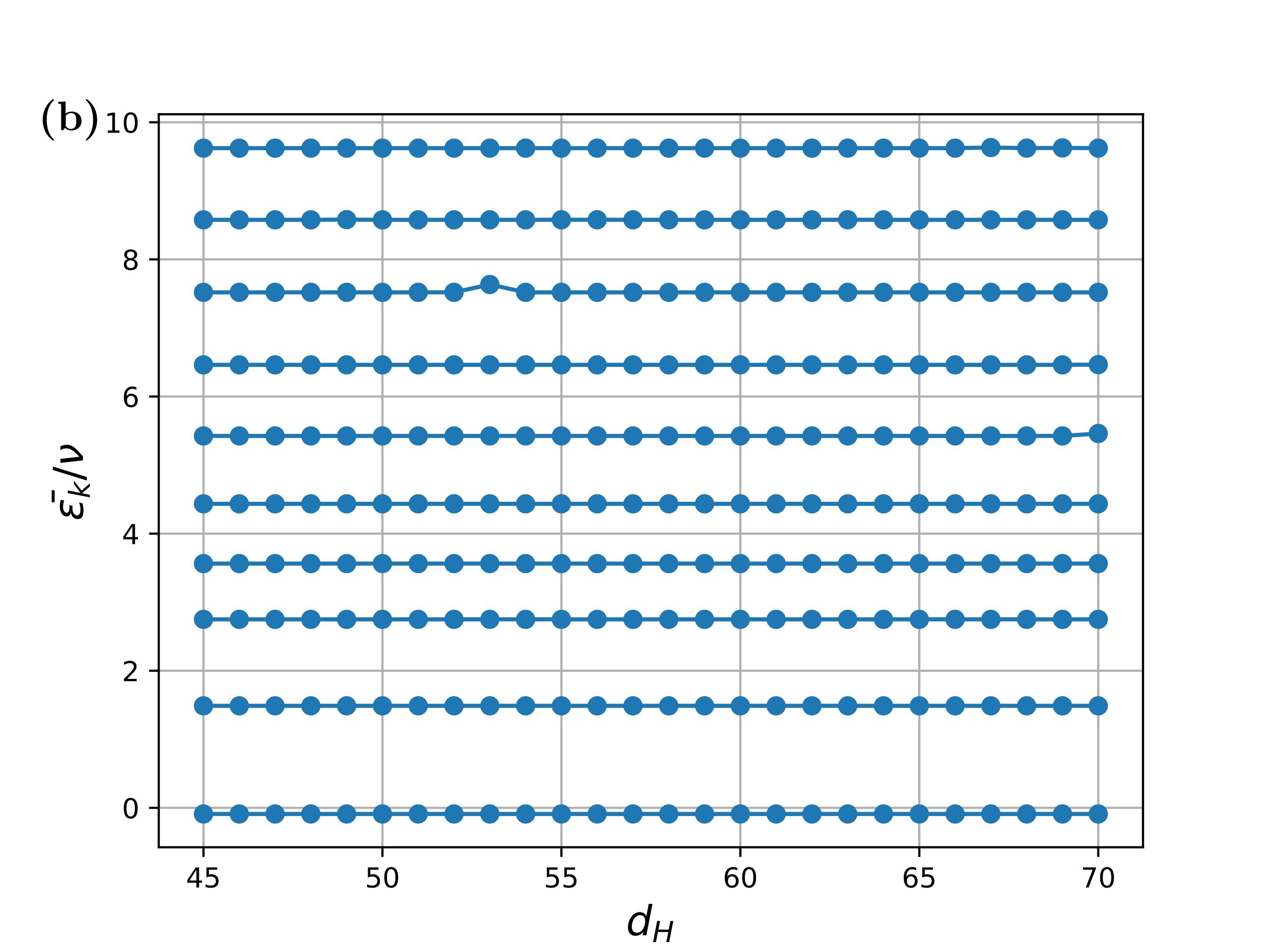}
    \end{subfigure}
    \caption{\textbf{Convergence of quasi-energies with Hilbert-space dimension.} The period averaged quasi-energies (a) at freezing, $\alpha=2$, and (b) away from freezing, $\alpha=1$, for various Hilbert space dimensions $d_H$ up to a maximum of $70$. The result is normalized by the fluxonium oscillator energy $\nu$ (\eqref{eq:osc_freq}). In both cases $\omega/\nu=12.31$, and all other parameters are as in Fig.~\ref{fig:IPR}.  
}\label{fig:Hilbert_space_dim}
\end{figure*}

As discussed in the main text, an important way that phase errors develop between the computational states is due to the dependence of the eigenenergies on $\vp_\text{ext}$. With this in mind, let us define
\begin{align}
    \Delta \epsilon_{01}(\vp_\text{ext}) = \frac{\epsilon_1(\vp_\text{ext}) - \epsilon_0(\vp_\text{ext})}{\nu}\label{eq:E_gap}
\end{align} to be the normalized energy gap between the computational states. This dimensionless quantity determines the rate at which phase errors develop at a given $\vp_\text{ext}$ for a system initialized at $\pi$-flux.

To determine this value analytically, we must go to second order in the Floquet-Magnus expansion. The second order terms are given by

\begin{align}
    H_\text{eff}^{(2)} =& -\frac{1}{6T}\int_0^T dt_1\int_0^{t_1}dt_2\int_0^{t_2}dt_3\\
    &\Big(
    [{\cal{H}}_{\tn{mov}}(t_1),[{\cal{H}}_{\tn{mov}}(t_2),{\cal{H}}_{\tn{mov}}(t_3)]]\nonumber\\
    &+ [{\cal{H}}_{\tn{mov}}(t_3), [{\cal{H}}_{\tn{mov}}(t_2),{\cal{H}}_{\tn{mov}}(t_1)]]\Big)\nonumber.
\end{align} At the $k$th freezing point for the triangular-wave drive, we find that $\Delta \epsilon_{01}(\vp_\text{ext})$ is given by the expression in \eqref{eq:E_mag_second_order}. In fact, exponential suppression observed likely extends to all energy differences between states. This is because the matrix elements of $\cos\hat\vp$ with respect to the bosonic basis have exponentially suppressed magnitude in terms of $\sqrt{8E_C/E_L}$ \cite{Catelani_2011}. We find that all the terms in the second order Magnus Hamiltonian have polynomial dependence on $\cos\hat\vp$ resulting in an exponential suppression at all orders in perturbation theory (as the corrections to the energy are polynomial in the matrix elements at every order in perturbation theory). This argument breaks down when the coefficient of anharmonic term is no longer perturbative with respect to the harmonic oscillator Hamiltonian. As such, this exponential suppression only occurs near freezing points where the anharmonic terms are suppressed by powers of $1/\omega$. The analytic form of higher differences $(\epsilon_m-\epsilon_{m-1})/\nu$ that we have computed share this exponential suppression supporting the above argument.

In Fig~\ref{fig:dispersion} we plot the normalized period-averaged quasi-energy gap, the second order prediction of the Magnus expansion, and the normalized quasi-energy gap (not period-averaged) up to modulo $\omega/\nu$ as a function of $\vp_\text{ext}$ for the driven system at the $\alpha = 2$ freezing point near $\pi$-flux. We find that the second order Magnus prediction agrees well with the normalized quasi-energy gap. Nonetheless, the quasi-energy gap and the period-averaged quasi-energy gap are found to differ. This difference stems from the effects of the Floquet micromotion operator \cite{Eckardt_2015} (which we have thus far ignored) which can be shown to affect the magnitude of period-averaged quasi-energies at orders $1/\omega^2$ and above. While we note that the dependence of period-averaged quasi-energy and quasi-energy on $\vp_\text{ext}$ are of similar order in our system, we leave a detailed high-order analysis of period-averaged quasi-energies to future work.\\

\section{Convergence with Hilbert Space Dimension}

In this brief section we discuss the convergence of the system simulations with Hilbert space dimension. In Fig.~\ref{fig:Hilbert_space_dim} we plot the period averaged quasi-energies for frozonium in the presence of a drive both at and away from freezing as a function of Hilbert space dimension. In both cases we find the quantities converge well for the lower lying states.

\bibliography{Reference.bib,references-vf.bib}

\end{document}